\documentclass[usenatbib]{mn2e}
\usepackage{graphicx,times}
\usepackage{bm}
\usepackage{epsf}
\usepackage{amsmath}
\usepackage{amssymb}
\usepackage{color}
\usepackage{enumerate}
\def\la{\langle}
\def\ra{\rangle}
\def\n{\noindent}
\def\be{\begin{equation}}
\def\ee{\end{equation}}
\def\ben{\begin{eqnarray}}
\def\een{\end{eqnarray}}
\def\nn{\nonumber}
\def\oh{\hat\Omega}

\def\bk{{\bf k}}

\def\br{{\bf r}}
\def\inc{{\int_0^{r_s}}}

\def\cI{{\cal I}}

\def\bk{{\bf k}}

\def\bl{{\bf l}}
\def\bx{{\bf x}}
\def\2p{{(2\pi)^2}}

\def\bl{{\bf l}}
\def\be{\begin{equation}}
\def\ee{\end{equation}}
\def\beq{\begin{equation}}
\def\eeq{\end{equation}}
\def\ben{\begin{eqnarray}}
\def\een{\end{eqnarray}}

\def\oh{{\hat\Omega}}
\def\nn{{\nonumber}}

\newcommand{\beqa}{\begin{eqnarray}}
\newcommand{\eeqa}{\end{eqnarray}}

\def\ex1{{\int {d^2 {\bf l} \over (2
\pi)^2}~ {\rm P}_{\pi} { \big ( {l\over d_A(r)} \big )} b_l^2(\theta_b)}}
\def\av{\langle  \tilde y_{s}^2(\theta_b) \rangle_c}
\def\kmin1{{\int_0^{r_s}\; \omega_{\rm SZ}(r)\; dr }}
\def\one{\langle  \tilde y_{s}^2(\theta_b) \rangle_c}
\def\two{\langle  \tilde y_{s}(\oh_1) \tilde y_{s}(\oh_2) \rangle_c}

\def\tI{{\cal I}_{\theta_b}}
\def\ex{{\cal I}_{\theta_b}}
\def\kmin{{\Theta_m}}

\def\cbr{\textcolor{black}}
\def\cb{\textcolor{black}}

\begin{document}
\onecolumn
\title[Statistical Properties of Thermal Sunyaev-Zel'dovich Maps]
{Statistical Properties of Thermal Sunyaev-Zel'dovich Maps}
\author[D.Munshi et al.]
{Dipak Munshi$^{1}$, Shahab Joudaki$^{2}$, Joseph Smidt$^{2}$, Peter Coles$^1$,
 Scott T. Kay$^3$  \\
$^{1}$School of Physics and Astronomy, Cardiff University, Queen's
Buildings, 5 The Parade, Cardiff, CF24 3AA, UK \\
$^{2}$ Department of Physics and Astronomy, University of California, Irvine, CA 92697\\
$^{3}$ Jodrell Bank Center for Astrophysics, Alan Turing Building, The University
of Manchester, M13 9PL, UK}
\maketitle
\begin{abstract}
On small angular scales, i.e. at high angular frequencies, beyond
the damping tail of the primary power spectrum, the dominant
contribution to the power spectrum of cosmic microwave background
(CMB) temperature fluctuations is the thermal Sunyaev-Zel'dovich
(tSZ) effect. We investigate various important statistical
properties of the Sunyaev-Zel'dovich maps, using well-motivated
models for dark matter clustering to construct statistical
descriptions of the tSZ effect to all orders enabling us to
determine the entire probability distribution function (PDF). Any
generic deterministic biasing scheme can be incorporated in our
analysis and the effects of projection, biasing and the underlying
density distribution can be analyzed separately and transparently in
this approach. We introduce the {\em cumulant correlators} as tools
to analyze tSZ catalogs and relate them to  corresponding
statistical descriptors of the underlying density distribution. The
statistics of hot spots in frequency-cleaned tSZ maps are also
developed in a self-consistent way to an arbitrary order, to obtain
results complementary to those found using the halo model. We also
consider different beam sizes, to check the extent to which the PDF
can be extracted from various observational configurations. The
formalism is presented with two specific models for underlying
matter clustering, the hierarchical ansatz and the lognormal
distribution. We find both models to be in very good agreement with
the simulation results, though the extension of hierarchical model
has an edge over the lognormal model. In addition to testing against
simulations made using semi-analytical techniques we have also used
the maps made using Millennium Gas Simulations to prove that the PDF
and bias can indeed be predicted with very high accuracy using these
models. The presence of significant non-gravitational effects such
as pre-heating, however, can not be modeled using an analytical
approach which is based on the modeling of gravitational clustering
alone. Our results indicate that the PDFs we construct are
insensitive to the underlying cosmology and can thus provide a
useful probe of non-gravitational processes e.g. pre-heating or
feedback.
\end{abstract}
\section{Introduction}
The inverse Compton scattering of CMB
photons - known as the thermal Sunyaev-Zel'dovich effect (tSZ;
\cite{SZ72,SZ80,Rep95,Bir99}) - imprints a characteristic distortion
in the Cosmic Microwave Background (CMB) spectrum that can be studied
using surveys such as WMAP\footnote{http://wmap.gsfc.nasa.gov/} and the ongoing
Planck\footnote{http://www.rssd.esa.int/Planck} satellite mission.
The fluctuation of this distortion
across the sky as probed by CMB observations can thus provide valuable
clues to the fluctuations of the gas density and temperature.  The
up-scattering in frequency of CMB photons implies an increment in
the spectrum at high frequencies with corresponding decrement in the
low frequency (Rayleigh-Jeans; RJ) regime, and a null around $217$
GHz. This characteristic behaviour is a potential tool for the
separation of tSZ from the other temperature anisotropy
contributions. These techniques are extremely effective in
subtraction of primary anisotropies due to its well understood
(perfect black body) frequency dependence and near Gaussian
statistical behaviour \citep{BG99,DCP03,Leach08}.
The tSZ effect is now routinely imaged in massive galaxy clusters
where the temperature of the scattering medium can reach as high as
$10{\rm\cb{keV}}$. This in effect produces a change in CMB temperature
of order $1$mK at RJ  wavelengths.

Here we are interested in the general
intergalactic medium (IGM) where the gas is expected to be at $\le
{\rm 1keV}$ in the sort of mild over-densities that lead to CMB
contributions in the $\mu$K range. In this work we primarily focus
on the statistical study of wide-field CMB data where tSZ effects
lead to anisotropies in the temperature distribution both due to
resolved and unresolved galaxy clusters, keeping in mind that the
thermal tSZ contribution is the dominant signal beyond the damping
tail of the primary anisotropy power spectrum. We primarily focus on
analytical modelling of the entire one and the joint two-point PDF of the
tSZ effect.

The tSZ power spectrum is known to be a sensitive probe of the
amplitude of density fluctuations. Higher order statistics such as the
skewness or bispectrum can provide independent estimates on the bias
associated with baryonic pressure, as well as providing further
consistency checks and cross-validation of lower order estimates.
The modelling of lower order statistics of the tSZ effect done by various
authors \citep{Sj00,AC1,Zp01,AC2,KS02,ZS07} in the past has followed
the halo model \citep{CooSeth02} that relies on
ingredients for the mass function based on the Press-Schechter \citep{PS74} formalism
and radial profile given by \cite{NFW96}.

In addition to analytical modelling, the numerical simulation of tSZ
plays an important role in our understanding of the physics involved
\citep{Persi95,dasilva99,Ref00,Sel01,Spr01,White02,Lin04, Z04,
Cao07,Ron07,Hal07,Hal09,Scott12}.Though limited by their dynamic range, some of these studies
incorporates complication from additional radiative and
hydrodynamical effects (i.e. ``gastrophysics'') such as radiative
cooling, preheating and SN/AGN feedback to a certain extent which are
otherwise difficult to incorporate in any analytical
calculations.

In parallel with the development of this PS formalism, analytical
modelling based on hierarchical form for the higher order correlation
functions has also been studied extensively \citep{SSZ92,szsz93,sz96,szsz97,Bernardreview02}.
We employ the particular form proposed
by \citep{BS89} to model the tSZ statistics. This form has been
studied extensively in the literature for modelling weak lensing as
well for the statistics of collapsed objects and related
astrophysical phenomenon. The statistics of collapsed objects and
their contribution to the tSZ sky have been studied previously
\citep{VS99,VSS02}. These studies also probe X-ray luminosity from
the same clusters. In this study we do not probe individual clusters
or collapsed objects, but instead directly link the density field
with corresponding SZ observables. Though the diffuse component of the
tSZ effect is beyond WMAP detection threshold the situation may improve with
future data sets such as Planck \citep{Han05,Joudaki10} or surveys such as
Arc-minute Cosmology Bolometer Array Receiver (ACBAR); see
\cite{Run03}\footnote{http://cosmology.berkeley.edu/group/swlh/acbar/}

The paper is organized as follows. In \textsection\ref{sec:for} we
provide the details of tSZ effect. We link the higher order
multispectra of the SZ effect with the underlying mass distribution
with the help of various biasing schemes in
\textsection\ref{sec:lower}. In \textsection\ref{sec:hier} we
introduce the generic hierarchical \emph{ansatz} and in
\textsection\ref{sec:gen} we introduce the specific formalism based
on generating functions in the quasi-linear and highly nonlinear
regime. In \textsection\ref{sec:pdf_bias} we show how the PDF and
bias of the tSZ sky are related to that of underlying density PDF
and bias. Various approximation schemes are discussed that can be
used to simplify the PDF and bias. In \textsection\ref{sec:sim} we describe
various simulations we have used in our study. In In \textsection\ref{sec:test}
we present the results of tests agains simulations. Finally the
\textsection\ref{sec:conclu} is dedicated to the discussion of our
result and future prospects.
\section{Formalism}
\label{sec:for}
In this section we will provide necessary theoretical background for
the computation of lower order moments of tSZ both for the one-point cumulants
and the two-point cumulant correlators. These will be later used to
construct the entire PDF  and the bias of tSZ
in the context of hierarchical clustering.
 We will be using the following form of the Robertson-Walker line element for the
background geometry of the universe:
\begin{equation}
ds^2 = -c^2 dt^2 + a^2(t)[ dr^2 + d_A^2(r)(d\theta^2 + \sin^2\theta d\phi^2)].
\end{equation}
Where we have denoted the comoving angular diameter distance by $d_A(r)$ and
scale factor of the universe by $a(t)$. $d_A(r)= {\rm K}^{-1/2}\sin
({\rm K}^{\cb{+}1/2} r)$ for positive curvature, $d_A(r) = (-{\rm
K})^{-1/2}\sinh ((-{\rm K})^{\cb{+}1/2}r)$ for negative curvature and $r$
for a flat universe. Here $r$ is {\em comoving distance or lookback time}. 
For a present value of  of ${\rm H}_0$ and
$\Omega_0$ we have ${\rm K}= (\Omega_0+\Omega_{\Lambda} -1){\rm H}_0^2$. The thermal
Sunyaev-Zel'dovich (tSZ) temperature fluctuation $\Delta T_{\rm SZ}(\oh,\nu)=\delta T(\oh,\nu)/T_{\rm CMB}$ is given by the opacity
weighted electron pressure:
\be
\Delta T_{\rm SZ}(\oh,\nu) \equiv g_\nu(x_{\nu})y(\oh) = g_\nu(x_{\nu})\int_0^{r_0}  dr \; \dot \tau_e \; \pi_e(\oh,r); \quad \pi_e({\bf x}) = \delta p_e({\bf x}) / \la p_e \ra.
\ee
\n Here $\tau_e$ is the Thomson optical
depth; overdots represent derivatives with respect to $r$. Here
$y(\oh)$ is the map of the Compton $y$-parameter. The Thomson optical depth  $\tau_e$
can be expressed in terms the
Thompson cross-section $\sigma_{\rm T}$, is the Boltzman constant
$k_{\rm B}$ by the integral $\tau_e = c\int n_e(z)\sigma_t dt$. The
free electron number density is represented by $n_e(z)$. The
function $g_\nu(x)$ encodes the frequency dependence of the tSZ
anisotropies. It relates the temperature fluctuations at a frequency
$\nu$ with the Compton parameter $y$. Here the function $g_\nu(x)$
is defined as: $g_\nu(x_{\nu}) = x_{\nu}\coth \left({x_{\nu}/2}\right) -4$ with $\quad
x_{\nu}= {h\nu /(k_B T_{\rm CMB})} = {\nu / (56.84 {\rm GHz})}$.  In the
low frequency part of the spectrum $g_\nu(x_{\nu}) = -2$, for $x_{\nu} \ll 1$,
here $x_{\nu}$ is the dimensionless frequency as defined above. We will
primarily be working in the Fourier domain and will be using the
following convention:
\ben
{\pi}_e(\bk;r) = \int d^3 \bx \;{\pi}_e(\bx;r) \exp[{-i\bk\cdot\bx}].
\een
\n The projected statistics that we will consider can be related to
3D statistics defined by the following expressions
which specify the power spectrum $P_{\pi}(k;r)$, the bispectrum
$B_{\pi}(k_1,k_2,k_3;r)$ and the trispectrum
$T_{\pi}(k_1,k_2,k_3,k_4;r)$ in terms of the Fourier coefficients. In
our notation we will separate the temporal dependence with a semicolon:
\ben
&& \la \pi_e(\bk_1;r)\pi_e(\bk_2;r) \ra_c = (2\pi)^3\delta_D(\bk_1 + \bk_2) P_{\pi}(k_1;r);~
\la \pi_e(\bk_1;r)\pi_e(\bk_2;r)\pi_e(\bk_3;r) \ra_c = (2\pi)^3 B_{\pi}(k_1,k_2,k_3;r)\delta_D({\bk_1 + \bk_2+\bk_3}) \quad\\
&& \la \pi_e(\bk_1;r)\pi_e(\bk_2;r)\pi_e(\bk_3;r)\pi_e(\bk_4;r) \ra_c = (2\pi)^3 T_{\pi}(k_1,k_2,k_3,k_4;r)\delta_D({\bk_1 + \bk_2+\bk_3+\bk_4}).
\label{eq:multi}
\een
\begin{figure}
\begin{center}
{\epsfxsize=7.95 cm \epsfysize=7.95 cm {\epsfbox[6 6 423 365]{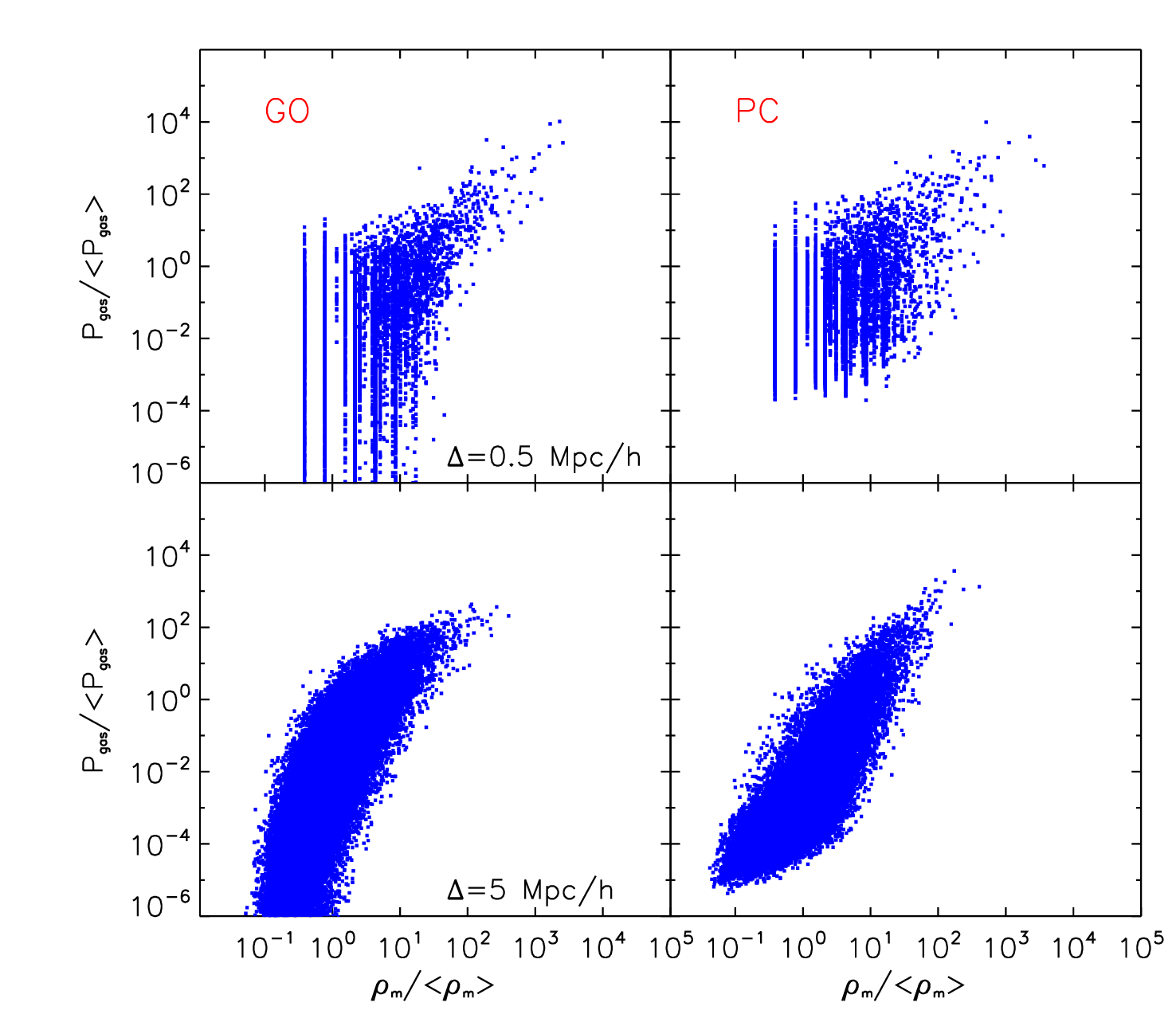}}}
\end{center}
\caption{Scatter plots of the baryonic pressure vs. the dark matter density contrast 
$1+\delta=\rho_m/\la\rho_m\ra$
are displayed for various simulations used in our study at a redshift $z=0$. The left panels correspond to the simulations with no preheating (i.e. gravity
only simulations but with adiabatic cooling), shown here as GO, where as
the right panels correspond to the simulations with preheating, depicted as PC. 
More details about the simulations are presented in \textsection\ref{sec:sim}. 
The smoothing scale for these plots is $0.5 h^{-1}{\rm Mpc}$ for the upper panels and
$5 h^{-1}{\rm Mpc}$ for the lower panels. 
The scatter is lower for regions with high density contrast  $1+\delta>100$ where
most of the tSZ signals originate.
The simulations with preheating exhibits a less well defined correlation structure.
We have shown 50,000 randomly sampled points from our simulations in each panel.}
\label{fig:scatter}
\end{figure}
\begin{table*}
\label{tab_notation}
\begin{center}
\begin{tabular}{|c |c| c|c|c|c|c }
\hline
\hline
 & cumu. & cor. & VPF & 2VPF & 1PDF & 2PDF \\
\hline
$\delta(r)$ & $S_p $ & $C_{pq}$ & $\phi(z)$ & $\beta(z)$ & $h(x)$ & $b(x)$ \\
$\tilde y_s(\oh)$ & $\tilde S_p $ & $\tilde C_{pq}$ & $\tilde \Phi(z)$ & $\tilde\beta(z)$ & $\tilde h(x)$& $\tilde b(x)$\\
$\hat y_s(\oh)$ & $\hat S_p $ & $\hat C_{pq}$ & $\hat\Phi(z)$ & $\hat\beta(z)$ & $\hat h(x)$ & $\hat b(x)$\\
\hline
\hline
\end{tabular}
\caption{The notations for various statistics of $\delta(r)$, $\tilde y_s(\oh)$ and $\hat y_s(\oh)$ are tabulated.
The parameter $\hat y$ is constructed to have same statistics as $\delta$ under certain simplifying approximation
i.e. $\hat h(x) =h(x)$ and $\hat b(x)=b(x)$. The variance of $\delta(r)$ is however different compared to that of
$\hat y_s(\oh)$. Also, notice that $\delta(r)$ is a 3D field whereas $\hat y(\oh)$ is a projected (or 2D) field. The normalised cumulants of  $\hat y(\oh)$ and $\delta(r)$ are identical, under
certain approximation. Hence they are independent of the biasing. The variance of these fields are
however different.}
\end{center}
\label{table:notation}
\end{table*}
\n
\cb{The subscript $c$ denotes the {\em connected} parts of a cumulant, i.e. those parts not related to the lower-order correlations}.
The multispectra of the underlying density field will  develop a
similar hierarchy. In the linear bias formalism, the multispectra
of the field $\pi$ are directly linked to the density contrast
$\delta = (\rho_m-\la\rho_m\ra)/\la\rho_m\ra$ with $\rho_m$ being the homogeneous
background density of the Universe. Using a bias factor
$b_{\pi}(r)$, which depends on redshift, we can write \citep{GS99a,GS99b,AC1,AC2}:
\ben
P_{\pi}(k_1;r) = b^2_{\pi}(r) P_{\delta}(k_1;r); \quad\quad
B_{\pi}(k_1,k_2,k_3;r_i) =  b^3_{\pi}(r) B_{\delta}(k_1,k_2,k_3;r_i); \quad\quad
T_{\pi}(k_1,k_2,k_3,k_4;r_i) =  b^4_{\pi}(r) T_{\delta}(k_1,k_2,k_3,k_4;r_i).
\label{eq:multi2}
\een
\n The redshift $z_s$ dependence of the bias is typically assumed to be of the
following form: $b_{\pi}(z_s) = {b_{\pi}(0)/(1+z_s)}$ and $b_{\pi}(0)=
k_B T_e(0)b_{\delta}/(m_ec^2)$. In our analysis we will show that
it is possible to define a reduced $y$ parameter whose statiscs 
will be insensitive to the details of the biasing model. The biasing scheme
is motivated by a tight correlation of baryonic pressure and the density
contarst at high density regions where most of the tSZ signals originate.

We have studied the correlation between the fractional baryonic gas pressure 
$\rm P_{gas}/ \la\rm P_{gas} \ra$ 
and the density contrast $\rho_m/\la \rho_m\ra = 1+\delta$ in
Figure \ref{fig:scatter}. The electronic pressure ${p_e}$ and 
the baryonic (gas) pressure ${\rm P_{gas}}$
are related \citep{AC1,Cooray01,AC2}  ${p_e} ={3{\rm X}+2 \over 3{\rm X}+5} {\rm P_{gas}}$; here ${\rm X}=0.76$ is the primordial
hydrogen abundance. The 3D electronic pressure fluctuation $\pi_e$ can be 
expressed as $1+\pi_e = {p_e}/\la { p_e}\ra = 
{\rm P_{gas}}/\la {\rm P_{gas}}\ra.$
We find a correlation between these two variables i.e. $\delta$ and $\pi_e$.
The correlation is tighter for regions with higher density contrast $\delta$,
where most of the tSZ signals originate. We have considered two
smothing scales. The top and bottom panels    
correspond to smoothing scales of $.5h^{-1}$Mpc and $5h^{-1}$Mpc respectively.
We have displayed 50,000 randomly selected points from the simulations 
used for our study at a redshift of $z=0$. The left panels correspond to
the gravity only or GO simulation where as the right panels correspond to
simulation with preheating and are denoted by PC. The correlation is more
pronounced for the GO simulations.
%
\section{Lower-Order Statistics of the Thermal SZ Effect}
\label{sec:lower}
The statistics of the smoothed tSZ effect $\hat y_{s}= {(y_s -\la y_s \ra)/ \la y \ra }$
reflect those of the baryonic pressure fluctuations projected along the line of
sight. Notice that in the denominator we have the average of unsmoothed $y$ parameter.
In our analysis we will consider a small patch of the sky
where we can use the plane parallel (or small angle)
approximation to replace spherical harmonics by Fourier modes. The
three-dimensional electronic pressure fluctuations $\pi_e(\bx)$ along the line of
sight when projected onto the sky with the weight function
$\omega_{\rm SZ}(r)$ give the tSZ effect in a direction $\oh$ which
we denote by $y_{}({\oh})$ (the smoothed y-maps will be
denoted by $y_{s}({\oh})$ where subscript $s$ will denote smoothed quantities):
\be y_{s}({\oh}) =
\inc {dr}\; \omega_{\rm SZ}(r)\;\pi_e(r,{\oh}); \quad
\omega_{\rm SZ}(r) = \dot \tau_e(r);\quad\quad y_s(\oh)=\int d\oh'{\rm W_G}(\oh-\oh';\theta_b)y(\oh').
\ee
Throughout, we will be using a Gaussian window ${\rm W_G}$, or equivalently
a Gaussian beam $b_l$ in the harmonic domain, specified by its
full width at half maxima or FWHM $=\theta_b$. 
Using the small angle approximation we can compute the projected two-point
correlation function in terms of the dark matter power spectrum
$P_{\delta}(k,r)$ \citep{NK92}:
%
%
\begin{equation}
\langle \tilde y_{s}(\oh_1) \tilde y_{s}(\oh_2) \rangle_c = \inc d {r}
{\omega_{\rm SZ}^2(r) \over d^2_A(r)} \int {d^2 {\bf l} \over (2
\pi)^2}~\exp ( \cb{i}\; {\bf \theta}_{12} \cdot {\bf l} )~b_l^2(\theta_b) {P}_{\pi} { \left [ {l\over d_A(r)}\; \cb{;} \; r \right ]}; \quad \tilde y_{s} = y_s - \la y_s \ra .
\end{equation}
\n Here $\theta_{12}$ is the angular separation projected onto the
surface of the sky and we have also introduced ${\bf l} = d_A(r){\bf
k}_{\perp}$ to denote the scaled projected wave vector. Using
Limber's approximation, the variance and higher order moments of $y_{s}$, smoothed using
a Gaussian beam: $b_l(\theta_b)= \exp[-l(l+1)\sigma_b]; \quad 
\sigma_b = {\theta_b \over \sqrt{8\ln(2)}}$ with FWHM = $\theta_b$, can be
written as:
\begin{equation}
\langle \tilde y_{s}^p({\theta_b}) \rangle_c = \inc d {r}
{\omega_{\rm SZ}^p(r) \over d_A^{2\cb{(p-1)}}(r)} \int {d^2 {\bf l_1} \over (2
\pi)^2} b_{l_1}(\theta_b) \cdots \int {d^2 {\bf l_{p-1}} \over (2
\pi)^2} b_{l_{p-1}}(\theta_b)~b_{l_p}(\theta_b)~ {B}^{(p)}_{\pi}
\cb{ \Big ( {{\bf l}_1\over d_A(r)},\cdots, {{\bf l}_p\over d_A(r)} , r \Big )}
\cb{\delta_D({\bf l}_1+\cdots+ {\bf l}_p)}.
\label{kappa_variance}
\end{equation}
The higher-order moments of the smoothed temperature field
relate $\langle \tilde y^p_{s} ({\theta_b}) \rangle_c$ to the
three-dimensional multi-spectra of the underlying pressure fluctuations
$B^{(p)}_{\pi}$ \citep{Hui99,MC00}.
%
%
\n
We will use these results to show that it is possible
to compute the complete probability distribution function
of $y_{s}$ from the underlying dark matter probability
distribution function. Details of the analytical results presented
here can be found in \citep{MCM1,MCM2,MCM3,MC00}.
A similar analysis for the higher order cumulant correlators \citep{szsz97,MCM3} of the smoothed SZ field
relating $\langle \tilde y_s^p (\oh_1) \tilde y_s^q (\oh_2) \rangle_c$
with multi-spectra of pressure fluctuation $B^{(p+q)}_{\pi}$
can be expressed as \citep{MC00}:
\begin{eqnarray}
\langle \tilde y_{s}^p(\oh_1) \tilde y_{s}^q(\oh_2) \rangle_c = &&
\int_0^{r_s} { \omega_{\rm SZ}^{\cb{p+q}} (r) \over d_A^{\cb{2(p+q-1)}}(r) } dr \int
 \frac{d^2{\bf l}_1}{(2\pi)^2}b_{l_1}(\theta_b) \cdots  \int  \frac{d^2{\bf
 l}_{p+q-1}}{(2\pi)^2} b_{l_{\cb{p+q-1}}}(\theta_b)b_{l_{\cb{p+q}}}(\theta_b)
\exp(i\;({{\bf l}_1}+\cdots+{{\bf l}_{{p}}})\cdot\theta_{12}) \nn \\
&& \times \;\cb{{\rm B}^{(p+q)}_{\pi}\left( {{\bf l}_{1} \over d_A(r)}, \cdots ,{{\bf l}_{p+q} \over d_A(r)} \right)}
\cb{\delta_D({\bf l}_1+\cdots+ {\bf l}_{p+q})}.
\end{eqnarray}
We will use and extend these results in this paper to show that it
is possible to compute the entire bias function $b(> \hat y_{s})$, i.e. the bias associated with those spots in the tSZ map
where $\hat y_{s}$ is above certain threshold, from the
statistics of underlying over-dense dark objects; this then acts as
a generating function for the cumulant correlators. Details of the
analytical results presented here can be found in  \citep{MC00}.
%
%
\section{Hierarchical {\em Ansatze}\;: \; The Minimal Tree Model and its Extension}
\label{sec:hier}
\begin{figure}
\begin{center}
{\epsfxsize=9 cm \epsfysize=7.5 cm
{\epsfbox[41 287 550 700]{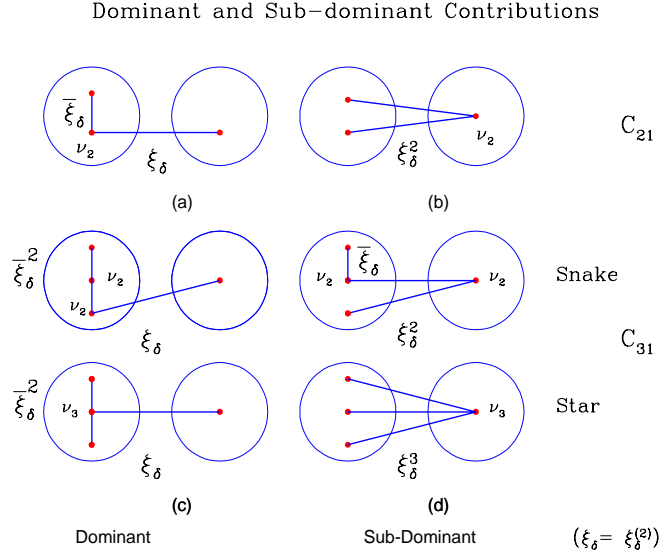}}}
\end{center}
\caption{A selection of diagrams that contribute to the cumulant correlators $C_{21}$ and $C_{31}$ are depicted.
At the lowest order in non-Gaussianity, the cumulant correlator $C_{21}$ has
two distinct contributions. The dominant contribution comes from the diagram (a).
The other contribution from the diagram (b) adds negligible contribution because
of the multiplicative factor $(\xi^{(2)}_\delta/\bar \xi^{(2)}_\delta)$. Here $\xi^{(2)}_\delta$ is the correlation function
and $\bar \xi^{(2)}_\delta$ is its average over the ``cell'' volume. We have suppressed the
superscripts for brevity. The two diagrams denoted by (c) shows
a representative dominant contribution to $C_{31}$. The diagrams denoted by (d) are
the sub-dominant contributions to $C_{31}$. The upper panels of diagram (c) and (d) are
of {\em snake} topology, where as the bottom panels correspond to the {\em star} topology.
In our analysis we have ignored the sub-dominant term which are negligible when the
cells are separated far apart.}
\label{fig:sim}
\end{figure}%
The spatial length scales corresponding to the small angular scales
of relevance here are in the highly non-linear regime. Assuming a
\emph{tree model }for the matter correlation hierarchy in the highly
non-linear regime for the density contrast $\delta$, one can write the general form of the $N$th order
correlation function $\xi^{\rm (N)}_{\delta}$ in terms of the two-point
correlation function $\xi^{(2)}_{\delta}$ \citep{Pee80,Fry84,B92}:
%
\begin{equation}
\xi^{(\rm N)}_{\delta}( {\bf r_1}, \dots {\bf r_N} ) = \sum_{\alpha, \rm \rm N-trees}
Q_{{\rm N},\alpha} \sum_{\rm labellings} \prod_{\rm edges}^{(\rm N-1)}
\xi^{(2)}_{\delta}({\bf r_i}, {\bf r_j}).
\end{equation}
It is interesting to note that a similar hierarchy develops in the
quasi-linear regime in the limit of vanishing variance \citep{BeS92}.
However the hierarchical amplitudes $Q_{\rm N, \alpha}$ become
shape-dependent functions in the quasi-linear regime. In the highly
nonlinear regime there are some indications that these functions
become independent of shape, as suggested by studies of the lowest
order parameter $Q_3 = Q$ using high resolution numerical
simulations \citep{SF99}. In Fourier space such an
\emph{ansatz} means that the hierarchy 
of multi-spectra can be
written as sums of products of the matter power-spectrum:
\begin{eqnarray}
&&B_\delta({\bf k}_1, {\bf k}_2, {\bf k}_3) = Q ( P_{\delta}({k_1})P_{\delta}({k_2}) + P_{\delta}({k_2})P_{\delta}({k_3})
+ P_{\delta}({k_3})P_{\delta}({k_1}) ); \\ \nonumber
&&T_\delta({\bf k}_1, {\bf k}_2, {\bf k}_3, {\bf k}_4) = R_a
\ P_{\delta}({k_1})P_{\delta}(|{\bk_1 +
\bk_2}|) P_{\delta}(|{ \bk_1 + \bk_2 + \bk_3}|)  + {\rm cyc. perm.} + R_b \ P_{\delta}({k_1})P_{\delta}({k_2})P_{\delta}({k_3}) +
{\rm cyc. perm.}
\end{eqnarray}
\n Different hierarchical models differ in the way they predict the
amplitudes of different tree topologies \citep{MCM1,MCM2,MCM3}.

Working directly with cumulant generating function $\phi(y)$ (to be 
introduced later), modeling of the entire probability distribution
function (PDF) was achieved by \citep{EPT97}. Remarkably, this was possible
by a simple empirical modification of the existing quasilinear 
predictions and was dubbed Extended Perturbation Theory (EPT).
The characteristic feature of this approach was to treat the local slope
of the linear power spectrum  as a free parameter in order to extend 
perturbative results to non-linear regime. EPT is fully parametrized 
by the non-linear variance and the skewness or equivalently $Q_3$ which
is predicted by Hyper Extended Perturbation Theory (HEPT) developed by \cite{SF99}.

In recent years a new semi-analytical model for PDF has been proposed by
\cite{VaMu04}\footnote{http://ipht.cea.fr/Pisp/patrick.valageas/codepdf\_en.php}.
Its construction is very similar to EPT and works directly with $\phi(y)$.
In addition to imposing the quasi-linear regime it also satisfies the
rare-void limit. We will use this approach and show that
the formalism is reasonably accurate in predicting the tSZ PDF.
\subsection{Cumulants}
\cb{Using this model we can express the one-point cumulants $\tilde S_N$ of $\tilde y_s = y_s - \la y_s \ra$ as:}
\ben
&& \langle \tilde y_{s}^3(\theta_b) \rangle_c = (3Q_3){\cal C}_3[\cI^2_{\theta_b}] \equiv S_3{\cal C}_3[\cI^2_{\theta_b}]
= \tilde S_3\langle \tilde y_{s}^2(\theta_b) \rangle_c^2 \label {hui}; \\
&& \langle \tilde y_{s}^4(\theta_b) \rangle_c = (12R_a + 4
R_b){\cal C}_4[\cI^3_{\theta_b}] \equiv S_4{\cal C}_4[\cI^3_{\theta_b}]= \tilde S_4^{} \langle \tilde y_{s}^2(\theta_b) \rangle_c^3.
\een
\n
In general, for an arbitrary order cumulant we can write:
\ben
\langle \tilde y_{s}^p(\theta_b) \rangle_c = S_p{\cal C}_{p}[[\cI_{\theta_b}]^{p-1}]; \quad
{\cal C}_p\left [[\cI_{\theta_b}]^{p-1}\right ] = \int_0^{r_s} { \omega_{\rm SZ}^p(r) b_{\pi}^{p}(r)
\over d_A^{2(p-1)}(r)}[\cI_{\theta_b}]^{(p-1)} dr; \quad\quad
[\cI_{\theta_b}] =  \int  \frac{d^2\bf l}{(2\pi)^2} P_{\delta}
\left( {l \over d_A(r)} \right) b_l^2(\theta_b).
\label{eq:SZ_sn}
\een
\n The projection effects on density cumulants were first derived in \cite{BWM97}.
In the context of weak lensing surveys 
the eq.({\ref {hui}}) was derived by \cite{Hui99}.
\cb{Notice that Eq.(\ref{eq:SZ_sn}) relates the normalised one-point cumulants of $ \tilde y_{s}$
i.e. $\tilde S_p$ to those of the underlying density contrast $S_p$.}

\cb{We will define a parameter $\hat y_s = \tilde y_s/\la y \ra$ which will further simplify
the results. In particular we will find that, under certain simplifying assumptions, the statistics of $\tilde y$\quad
are identical to those of the underlying density contrast $\delta$. Notice that the definition of the parameter $\hat y_s$ involves the unsmoothed $\la y_s \ra$
in the denominator. The normalised cumulants of $\hat y_s$ will be denoted as
$\hat S = \la \hat y_s^p \ra_c/\la \hat y_s^2 \ra_c^{(p-1)}$,
which can be expressed as $\hat S= \tilde S \la y \ra^{(p-1)}$.}
%
%
\subsection{Cumulant Correlators}
The concept of cumulant correlators was introduced by \cite{B96}.
Later studies extended this result to observational studies including weak-lensing 
statistics \citep{MC00,MJ00,MJ01}. 
\cb{In the present context, the family of cumulant correlators is important in modelling
the bias associated with $y$ maps.}
%
\begin{eqnarray}
\langle \tilde y_{s}^2(\oh_1) \tilde y_{s}(\oh_2) \rangle_c & = &
2Q_3 {\cal C}_3 [\cI_{\theta_b} \cI_{\theta_{12}}] \equiv C_{21}{\cal C}_3 [\cI_{\theta_b} \cI_{\theta_{12}}]
\equiv \tilde C_{21} \langle
\tilde y_{s}^2 \rangle_c \langle \tilde y_{s}(\oh_1) \tilde y_{s}(\oh_2) \rangle_c, \\
\langle \tilde y_{s}^3(\oh_1) \tilde y_{s}( \oh_2) \rangle_c & = &
(3R_a + 6 R_b){\cal C}_4 [\cI_{\theta_b}^2 \cI_{\theta_{12}}]
 \equiv  \tilde C_{31}^{} \langle
 \tilde y_{s}^2 \rangle_c^2 \langle \tilde y_{s}(\oh_1) \tilde y_{s}(\oh_2) \rangle_c.
\end{eqnarray}
In general, for an arbitrary order cumulant correlator we can write:
\be
\cb{\langle \tilde y_{s}^p(\oh_1) \tilde y_{s}^q(\oh_2) \rangle_c \equiv
C_{pq}^{}{\cal C}_{p+q} [[\cI_{\theta_b}]^{p+q-2} \cI_{\theta_{12}}] =
\tilde C_{pq}^{} \langle
\tilde y_{s}^2\rangle_c^{(p+q-2)} \langle \tilde y_{s}(\oh_1) \tilde y_{s}(\oh_2) \rangle_c,}
\ee
\n
where we have introduced the following notation:
\begin{equation}
{\cal C}_{p+q}[[\cI_{\theta_b}]^{(p+q-2)} \cI_{\theta_{12}} ] =
\int_0^{r_s} { \omega_{\rm SZ}^{p+q}(r) b^{p+q}_{\cb{\pi}}(r)\over
d_A^{2(p+q-1)}(r)}[{\cI}_{\theta_b}]^{p+q-2}[{\cI_{\theta_{12}}}] dr; \quad\quad
[\cI_{\theta_{12}}] = \int
 \frac{d^2\bf l}{(2\pi)^2} P_{\delta} \left( {l \over d_A(r)} \right)
b_l^2(\theta_b) \exp ({\cb i} {\bf l} \cdot {\bf \theta_{12}}).
\label{eq:cpq}
\end{equation}
\n
\cb{Notice that in the limiting case of  $\theta_{12}=0$ we recover the limiting
situation $\tilde C_{pq}^{}=\tilde S_{p+q}$ since we have $[\cI_{\theta_b}] =\cI_{\theta_{12}}$.
The cumulant correlators for $\hat y$
denoted as $\hat C_{pq}$ and $\tilde y$ are related by the expression : $\hat C_{pq} = \tilde C_{pq}\la y\ra^{p+q-1}$.}

These lowest order statistics can be helpful in probing the pressure bias
as a function of scale. It is however expected that the signal to noise will decrease
with increasing order. We will use these results to construct the entire
PDF and the bias associated with the tSZ maps. This will be achieved using
a generating function approach.
\begin{figure}
\begin{center}
{\epsfxsize=10 cm \epsfysize=5 cm
{\epsfbox[27 426 590 709]{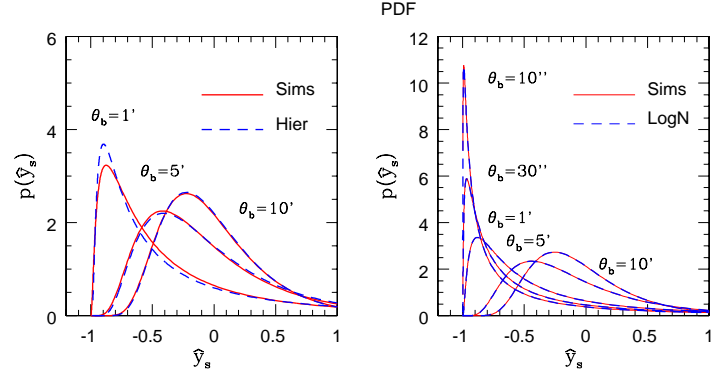}}}
\end{center}
\caption{The PDF $p(\hat y_s)$ is plotted as a function of $\hat
y_s$. Various curves correspond to different beam size (FWHM) as
indicated. Two different analytical models are shown; the lognormal
(right-panel) and the hierarchical ansatz (left-panel). The solid
lines correspond to the results from numerical simulations. The
dashed lines in each panel represent the analytical results.}
\label{fig:pdf}
\end{figure}
\section{The Generating Function}
\label{sec:gen}
To go beyond order-by-order approach discussed so far we will use
a formalism based on generating functions, which relies on the
hierarchical scaling nature of the higher order correlation function.
The scaling analysis deals directly with the generating functions
that encode the information regarding the correlation hierarchy. The
knowledge of these generating function are next useful in constructing
the one- and two-point PDFs.

In scaling analysis of the probability distribution function (PDF)
the Void Probability distribution function (VPF) plays a most
fundamental role, because it can be related to the generating
function $\phi(z)$ of the cumulants  or, if preferred, the $S_p$
parameters \citep{W79,BS89}:
\begin{equation}
P_v(0) = \exp \Big ( - { \phi (N_c)
\over \bar \xi_\delta^{(2)}} \Big  ),
\end{equation}
\n where $P_v(0)$ is the probability of having no ``particles'' in a
cell of of volume $v$, $\bar N$ is the average occupancy of these
``particles'' and $N_c = \bar N \bar {\xi_{}}^{(2)}_{\delta}$ and $\bar {\xi_{}}^{(2)}_{\delta}$
is the volume average of the two-point correlation function $\xi_{\delta}^{(2)}$.
Strictly speaking the above expression neglects any residual (subleading) 
$\bar{\xi}^{(2)}_\delta$ dependence of the $S_p$ parameters.
The VPF is meaningful only for discrete distribution of particles and can not be defined
for smooth density fields such as $\delta$ or $\hat y_{s}(\oh)$.
However the scaling function $\phi(z)$ defined above
 are very much useful even for continuous
distributions where they can be used as a generating function of
one-point cumulants or $S_p$ parameters: $\phi(z) =
\sum_{p=1}^{\infty} { S_p z^p /p! }$. The function $\phi(z)$
satisfies the constraint $S_1 = S_2 = 1$ necessary for proper
normalization of the PDF. The other generating function which plays
a very important role in such analysis is the generating function
for vertex amplitudes $\nu_n$, associated with nodes appearing in a
``tree'' representation of higher order correlation hierarchy ($Q_3 =
\nu_2$, $R_a = \nu_2^2$ and $R_b = \nu_3$).  In practice it is
possible to work with a perturbative expansion of the vertex
generating function ${\cal G}(\tau)$. In terms of the vertices it is
defined as: ${\cal G}(\tau) =  \sum_{n=0}^{\infty} (-\cb{\tau})^{n} { \nu_n/
n! }$. However in the highly nonlinear regime a closed form is used.
A more specific model for ${\cal G}(\tau)$, which is useful to make
more specific predictions (Bernardeau \& Schaeffer 1992) is given by
${\cal G}(\tau) = \Big ( 1 + {\tau / \kappa_a} \Big )^{-\kappa_a}$;
%
\be
\phi(z) = z {\cal G}(\tau) - { 1 \over 2} z {\tau} { d \over d
\tau} {\cal G}(\tau); \quad\quad \tau = -z { d \over d\tau} {\cal G}(\tau).
\label{eq:dummy_gen}
\ee
The range of $\delta$ for which the power law regime is valid depends on
the value of $\bar\xi_\delta^{(2)}$. For smaller values of $\bar\xi_2$ the power law
regime is less pronounced. A more detailed discussion of these issues can be
found in \cite{MunBer99}. The links to the gravitational dynamics in the quasilinear regime,
for various approximations are discussed in \cite{MSS94}.
\cb{The second equation of Eq.(\ref{eq:dummy_gen}) defines the variable $z$ in terms of $\tau$; which
can then be used to construct the function $\phi(z)$ for a given model of ${\cal G}(\tau)$.}
However a more detailed analysis is needed to include the effect of
correlation between two or more correlated volume elements which
will provide information about bias, cumulants and cumulant
correlators of these collapsed object (as opposed to the cumulants
and cumulant correlators of the whole  map, e.g. \cite{MJ00,MJ01}).
We will only quote results useful for measurement of bias;
detailed derivations of related results including related error
analysis can be found elsewhere \citep{BS99,MCM1,MCM2,CMM99,MCM3}

Notice that $\tau(z)$
(also denoted by $\beta(z)$ in the literature) plays the role of a
generating function for factorized cumulant correlators $C_{p1}$
($C_{pq} = C_{p1}C_{q1}$): $\tau(z) = \sum_{p=1}^{\infty}
{C_{p1}/p!} z^p$.
The PDF $p(\delta)$ and bias $b(\delta)$  can be related to their
generating functions VPF $\phi(z)$ and $\tau(z)$ respectively
by following equations \citep{BS89,BeS92,BS99}:
%
\be
p(\delta) = \int_{-i\infty}^{i\infty} { d{z} \over 2 \pi i} \exp \Big [ {(
1 + \delta )z - \phi({z})  \over \bar \xi^{(2)}_\delta} \Big ]; \quad
b(\delta) p(\delta) = \int_{-i\infty}^{i\infty} { dz \over 2 \pi i} \tau(z) \exp \Big [ {(
1 + \delta )z - \phi(z)  \over \bar \xi
^{(2)}_\delta} \Big ]
\label{eq:ber}. 
\ee\footnote{Eq.(\ref{eq:dummy_gen}) and Eq.(\ref{eq:ber}) are crucial
to any approach that uses the hierarchical ansatz for modeling of astrophysical
phenomena including the results presented here. 
These equations were derived by \cite{BS89} as well as \cite{BeS92,BS99}.}
\n It is clear that the function $\phi(z)$ completely determines the
behaviour of the PDF $p(\delta)$ for all values of $\delta$. The PDF too can be expressed with
the help of a scaling functions:
\be
p(\delta) = {h(x) \over [\bar \xi^{(2)}_\delta]^2}; \quad \quad \quad
x = {(1 + \delta) \over \bar \xi^{(2)}_{\delta}}.
\label{eq:pdf_scaling}
\ee
The scaling function
$h(x)$ is related to $\phi(z)$ through an inverse Laplace transform:
\be
h(x) = -{1 \over 2\pi i}\int_{-i\infty}^{+i\infty} {dz} \exp(x z) \phi(z); \quad\quad
h(x)b(x) = -{1 \over 2\pi i}\int_{-i\infty}^{+i\infty} {dz} \exp(x z) \beta(z).
\ee
The quantities $\phi(z)$ and $\beta(z)$ correspond to the density contrast $\delta$
and $\hat\Phi_{}(z)$ and $\hat\beta_{}(z)$ will denote corresponding quantities
for $\hat y_s$. The functions $b(x)$ is simply a scaled version of $b(\delta)$:
\be
b(x) = b\left ({1+\delta \over \bar \xi_\delta^{(2)}} \right ).
\ee
We will next use these expressions to analyse the PDF and bias of $y(\oh)$ maps.
\section{ The PDF and the bias of Beam Smoothed SZ Effect}
\label{sec:pdf_bias}
For computing the probability distribution function of the smoothed
scaled tSZ field we will use the variable $\tilde y_s$ that will make the analysis simpler: $\tilde y_{s}(\theta_b)= y_s(\theta_b) - \la y_s(\theta_b) \ra$, we will begin by
constructing its associated cumulant generating function
$\tilde \Phi(z)$:

\begin{equation}
\tilde \Phi_{{}}(z) = \sum_{p=1}^{\infty}\tilde S_p { z^p \over p!}  =  \sum_{p=1}^ {\infty} {{\langle
 \tilde y_{s}^p(\theta_b) \rangle_c} \over \langle \tilde y_{s}^2 (\theta_b )
\rangle_c^{p-1}}{ z^p \over p!}.
\end{equation}
\n Notice that the construction already satisfies the constraint
equation $S_1=S_2=1$. Now using the expressions for the higher
moments of the SZ in terms of the matter power spectrum
(see Eq.(\ref{eq:SZ_sn})) gives:
\begin{equation}
\tilde \Phi_{}(z) \equiv  \int_0^{r_s} \sum_{p=1}^{\infty}
{ 1 \over p!} S_p^{} \left [ { b^p_{\pi}(r)\omega_{\rm SZ}^p(r) \over d_A(r)^{2(p-1)} (r)}
{\tI^{ (p-1)} z^p \over \av^{(p-1)} } \right ];
\end{equation}
\n
We can now use the definition of $\phi(z)$ for the matter cumulants to
express $\tilde \Phi(z)$:
\begin{equation}
\tilde \Phi_{}(z) =  \int_0^{r_s} dr
\Big[ { d_A^2(r)\langle \tilde y^2(\theta_b)  \rangle_c  \over  \ex} \Big
] \phi \Big [ b_{\pi}(r) {\omega_{\rm SZ} (r) \over d_A^2 (r)} {\ex \over \av} z \Big ].
\end{equation}
Note that we have used the fully non-linear generating function $\phi$
for the cumulants, though we will use it to construct a generating
function in the quasi-linear regime.
\n
Next we relate these results to the statistics of previously defined quantity $\hat y_{s}$.
%
%
\n
For the reduced tSZ field $\hat y_s$, the cumulant generating function $\hat \Phi(z)$ is given by,
\begin{equation}
\hat \Phi_{} (z) = \sum_{p=1}^{\infty}\hat S_p { z^p \over p!} =
{1 \over {\la y(\theta_b)} \ra} \int_0^{r_s} dr \Big [{ d_A^2(r) \over \la y(\theta_b) \ra_c}{ \av \over \ex }\Big ]
\phi \Big[ b_{\pi}(r)\la y(\theta_b)\ra {z \over \la \tilde y_{s}^2(\theta_b) \ra_c} {w_{\rm SZ}(r) \over d_A^2(r)}{\mathcal I}_{\theta_b}\Big ].
\end{equation}
The scaling function $\hat h_{}(x)$ for $\hat y$ associated with the PDF $p_{}(\hat y)$ can now be
related with the matter scaling function $h(x)$ using the following definition \citep{BS89}:
\begin{equation}
\hat h_{} (x) = - \int_{-\infty}^{\infty} { dz \over 2 \pi i} \exp (x
z) \hat \Phi_{} (z);
\end{equation}
\n
which takes the following form, \cb{using definitions corresponding to Eq.(\ref{eq:pdf_scaling}),}
\be
\hat h_{}(x) = \int_0^{r_s} dr {w_{\rm SZ}(r) b_{\pi}(r) \over \la y(\theta_b) \ra}
\left [ {d_A^2(r) \over \la y(\theta_b)\ra }
{\la \tilde y_{s}^2({\theta_b})\ra_c \over \tI w_{\rm SZ}(r)b_{\pi}(r)} \right ]^2
h\left [x { d_A^2(r) \over {\mathcal I}_{\theta_b} w_{\rm SZ}(r)b_{\pi}(r)}
{\la \tilde y^2_{s}(\theta_b) \ra_c \over \la y(\theta_b)\ra} \right].
\ee
While the expressions derived above are exact, and are derived for
the most general case using only the small angle approximation, they
can be simplified considerably using further approximations. In the
following we will assume that the contribution to the $r$ integrals
can be replaced by an average value coming from the maximum of
$\omega_{\rm SZ}(r)$, i.e. $r_c$ ($0<r_c<r_s$). So we replace $\int
f(r) dr$ by $1/2 f(r_c)\Delta_{r}$ where $\Delta_{r}$ is the
interval of integration, and $f(r)$ is the function of comoving
radial distance $r$ under consideration.  Similarly we replace the
$\omega_{\rm SZ}(r)$ dependence in the ${\bf l}$ integrals by $\omega_{\rm
SZ}(r_c)$. Under these approximations we can write:
\be
\hat\Phi_{}(z) = \phi(z); \quad  \hat h_{}(x) = h(x).
\label{eq:scaled_vpf}
\ee
Thus we find that the statistics of the
underlying field $1+\delta$ and the statistics of the reduced field
$\delta y$ are exactly the same under such an approximation (the
approximate functions $\hat \Phi(z)$ and $\hat h(x)$ do satisfy the
proper normalization constraints). Although it is possible to
integrate the exact expressions of the scaling functions, there is
some uncertainty involved in the actual determination of these
functions and associated parameters that describes it
from N-body simulations (e.g. see Munshi et al. 1999, Valageas et
al. 1999 and  Colombi et al. 1996 for a detailed description of the
effect of the finite volume correction involved in their
estimation).

In the following, we use $\hat \Phi(z)$ as derived above
to compute $p(\delta \hat y)$ with the help of Eq.(\ref{eq:ber}).
In addition to the generating function approach we have used the lognormal
distribution as a model for the underlying statistics (see Appendix-A for a
detailed discussion on the lognormal distribution).
\subsection{The bias associated with the tSZ sky}
\label{sec:bias12}
\cb{The bias for the underlying density field $\delta$ is defined using the
joint two-point PDF $p(\delta_1,\delta_2)$ for density contrasts $\delta_1$ and $\delta_2$
measured at two different points separated by a fixed distance. Given the
two-point correlation function $\xi_{12}$ that characterizes the correlation
hierarchy for this scale, the 2PDF $p(\delta_1,\delta_2)$ can be expressed
in terms of the one-point PDFs and the bias functions $b(\delta)$ as follows:
\be
p(\delta_1,\delta_2)= p(\delta_1)p(\delta_2) \left [ 1+ b(\delta_1)\xi_{12} b(\delta_2) \right ].
\ee
The bias function $b(\delta)$ is also useful in describing the bias
associated with the over-dense regions. Cumulant correlators are
the lower-order connected moments of the 2PDF $p(\delta_1,\delta_2)$.}

To compute the bias associated with the peaks in the SZ field we
have to first develop an analytic expression for the generating field
$\tilde \beta(z_1, z_2)$ for the SZ field $\tilde y_{s}(\theta_b) = y_{s}(\theta_b)-
\la y(\theta_b)\ra $. For that we will use the usual
definition for the two-point cumulant correlator $C_{pq}$ for the
field:
\begin{equation}
\tilde C_{pq} =
{\langle \tilde y_{s}(\oh_1)^p  \tilde y_{s}(\oh_2)^q \rangle_c \over
\langle \tilde y_{s}^2(\theta_b) \rangle_c^{p+q-2} \langle
 \tilde y_{s}(\oh_1) \tilde y_{s}(\oh_2) \rangle_c };
\end{equation}
\n
for a complete treatment of two-point statistical properties of
smoothed fields see Munshi \& Coles (1999b).
We will show that, as is the case with its  density field
counterpart, the two-point generating function for the field
$y_{s}$ can also be expressed (under certain simplifying
assumptions) as a product of two one-point generating functions,
$\beta^{\rm SZ}(z)$, which can then be directly related to the bias
associated with ``hot-spots''in $y$-maps:
\begin{equation}
\hat\beta(z_1,z_2) =  \sum_{p,q}^{\infty} {\hat C_{pq} \over p! q!} z_1^p z_2^q =
\sum_{p}^{\infty} {\hat C_{p1} \over p!} z_1^p \sum_{q}^{\infty} {\hat C^{}_{q1}
\over q!} z_2^q  = \hat\beta_{}(z_1) \hat\beta_{}(y_2)\equiv
\hat\tau_{}(z_1) \hat\tau _{}(z_2).
\end{equation}
\n The tree-structure of the correlation hierarchy
that we have assumed is crucial to achieve the factorization derived above. 
It is also clear that the factorization of generating function
actually depends on the factorization property of the cumulant
correlators i.e. $\hat C^{}_{pq} = \hat C^{}_{p1}
\hat C^{}_{q1}$. Note that such a factorization is possible
when the correlation of two patches in the directions $\oh_1$ and
$\oh_2$ $\two$  is smaller compared to the variance $\one$ for the
smoothed patches. The generating function $\tilde \beta_{}(z_1,z_2)$ for $\tilde C_{pq}$ is constructed as follows: 
\begin{equation}
\tilde \beta_{}(z_1,z_2) = \sum_{p,q}^{\infty} {\tilde C_{pq} \over p! q!} { z_1^p z_2^q} \equiv
\sum_{p,q}^{\infty} {1 \over p! q!} { z_1^p z_2^q\over
\langle  \tilde y_{s}^2(\theta_b) \rangle_c^{p+q-2} } {\langle  \tilde y_{s}(\oh_1)^p  \tilde y_{s}(\oh_2)^q \rangle_c  \over \langle
\tilde y_{s}(\oh_1)\tilde y_{s}(\oh_2)\rangle_c }.
\end{equation}
\n We will now use the integral expression for the cumulant
correlators (Munshi \& Coles 1999a) in order to express the
generating function which, in turn, uses the hierarchical {\em
ansatz} and the far-field approximation as explained above. Using Eq.(\ref{eq:cpq}) we can write:
\ben
&& \tilde \beta_{}(z_1, z_2) = \sum_{p,q}^{\infty} { \tilde C^{}_{pq} \over p! q! } { z_1^p \over
\langle \tilde y_{s}^2(\theta_b) \rangle_c^{p-1}}{ z_2^q \over \langle y_{s}^2(\theta_b) \rangle_c^{q-1} } { 1 \over \tilde \xi_{12}}  \int_0^{r_s}\; dr\; d_A^2(r) b^{p+q}_{\pi}(r)
{\omega_{\rm SZ}^{p}(r) \omega_{\rm SZ}^{q}(r) \over d_A(r)^{2p} d_A(r)^{2q} } [{\mathcal I}_{\theta_b}]^{p+q-1}{\cal I}_{\theta_{12}}; \;\;\;  \\
&& \tilde\xi_{12} = \la \tilde y_s(\oh_1)\tilde y_s(\oh_2) \ra.
\een
\begin{figure}
\begin{center}
{\epsfxsize=10 cm \epsfysize=5 cm
{\epsfbox[27 426 590 709]{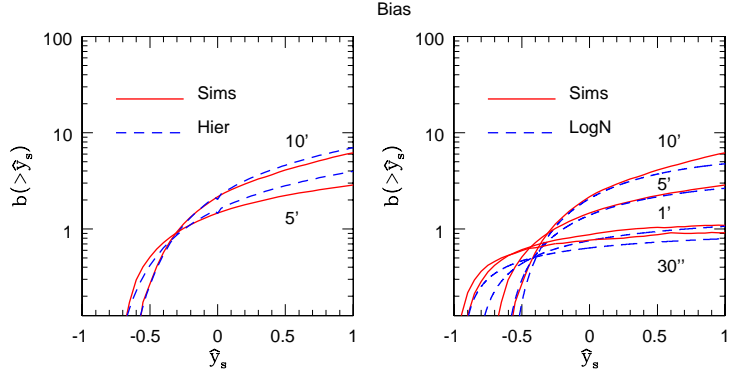}}}
\end{center}
\caption{The cumulative bias associated with $b(> \hat y_s)$ is
plotted as a function of $\hat y_s$. Various curves correspond to
different beam size (FWHM) as indicated. Two different analytical
models are shown; the lognormal (right-panel) and the hierarchical
ansatz (left-panel). the solid lines correspond to the results from
numerical simulations. The dashed lines in each panel represent the
analytical results.} \label{fig:bias}
\end{figure}
\n
It is possible to further simplify the above expression by separating the
summation over dummy variables $p$ and $q$, which will be useful to
establish the factorization property of two-point generating function
for bias $\hat\beta(z_1, z_2)$.
We can now decompose the double sum over the two indices into two
separate sums over individual indices.
\n The above expression is quite general and depends only on the
small angle approximation and the large separation approximation and
is valid for any given specific model for the generating function
${\cal G}(\tau)$. However it is easy to notice that the projection
effects as encoded in the line of sight integration do not allow us
to write down the two-point generating function
$\hat\beta(z_1, z_2)$ simply as a product of two
one-point generating functions $\hat\beta_{}(z)$ as was the
case for the density field $1+ \delta$.
As in the case of the derivation of the probability distribution
function it simplifies
matters if we use the reduced smoothed tSZ field $\hat y_s$. The
statistical properties of $\hat y_s$ are very similar to that of the
underlying 3D density field (under certain simplifying
approximations) and are roughly independent of the background
geometry and dynamics of the universe,
\be
\hat \beta(z_1,z_2) = \inc dr {1 \over \la y(\theta_b) \ra^2_{}} \; d^2_A(r)\;
{{\mathcal I}_{\theta_{12}} \over \tilde \xi_{12}}
{ \la \tilde y_s^2(\theta_b)\ra_c \over {{\mathcal I}_{\theta_b}}}
\hat \beta \Big ( b_{\pi}(r)\la y\ra { z_1 \over \la \tilde y_s^2(\theta_b) \ra_c} {\omega_{\rm SZ}(r) \over d^2_A(r)}{\mathcal I}_{\theta_b}  \Big )
{ \la \hat y_{s}^2(\theta_b) \ra_c
\over {{\mathcal I}_{\theta_b}}} \hat\beta \Big (b_{\pi}(r)\la y\ra { z_2 \over \la  \tilde y_{s}^2(\theta_b) \ra_c} {\omega_{\rm SZ}(r) \over d_A^2
(r)} {\mathcal I}_{\theta_b}  \Big ).
\ee
While the above expression is indeed very accurate and relates the
generating function of the density field with that of the tSZ field,
it is difficult to handle in practice. Also it is
important to notice that the scaling functions such as $h(x)$ for
the density probability distribution function and $b(x)$ for the
bias associated with over-dense objects are typically estimated from
numerical simulations specially in the highly non-linear regime.
Such estimations are plagued with several uncertainties such as
finite size of the simulation box. It was noted in earlier studies
that such uncertainties lead to only a rather approximate estimation
of $h(x)$. The estimation of the scaling function associated with
the bias i.e. $b(x)$ (here $x={(1+\delta)/\bar \xi_\delta^{(2)}}$) is even more complicated due to the fact that
the two-point quantities such as the cumulant correlators and the
bias are more affected by finite size of the catalogs. So it is not
fruitful to actually integrate the exact integral expression we have
derived above and we will replace all line of sight integrals with
its approximate values. \cb{Following our
construction of the one-point PDF or $p(\delta)$ we will replace integrals such as $\int_0^{r_s} f(r)dr$
with their approximate values $\int_0^{r_s} f(r)dr = {1/2}f(r_c)\Delta r$; $r_c$ is at an
intermediate redshift along the line-of-sight, its exact value is not important as the final
result will be independent of $r_c$. For more accurate
results we can Taylor expand $f(r)$ or integrate the exact expression:}
\begin{eqnarray}
&&\la y_{s}(\theta_b)\ra \approx {1\over 2} r_s b_{\pi}(r_c)\omega_{\rm SZ}(r_c),
\quad\quad \one  \approx {1\over 2} r_s  {\omega_{\rm SZ}^2(r_c) \over d^2_A(r_c)}b^2_{\pi}(r_c) \Big [ {\cb \int} {d^2 \bl \over
(2\pi)^2} {\rm P}_{\delta}\left ({l \over d_A(r_c)}\right ) b_l^2(\theta_b) \Big ],\\
&&\two  \approx {1\over 2} r_s  {\omega_{\rm SZ}^2(r_c) \over d^2(r_c)}b_{\pi}^2(r_c) \Big [ {\cb \int} {d^2 l \over
(2\pi)^2} {\rm P}_{\delta}\left ({l \over d_A(r_c)} \right ) b_l^2(\theta_b) \exp [i \; {\cb{\bf l}} \cdot \theta_{12}] \Big ].
\end{eqnarray}
\n Use of these approximations gives us the leading order
contributions to these integrals and we can check that to this order
we recover the factorization property of the generating function
i.e. $\hat \beta_{}(z_1,z_2) = \hat \beta_{}(z_1) \hat \beta_{}(z_2) =
\beta_{}(z_1) \beta_{}(z_2) \equiv
\hat \tau_{}(z_1)\hat \tau_{}(z_2)$. So it is clear that at this level of
approximation, due to the factorization property of the cumulant
correlators, the bias function $\hat b(x)$ associated with the
peaks in the field $\hat y_{s}$,  beyond certain threshold,
obeys a similar factorization property too, which is exactly same as
its density field counterpart. Earlier studies have established such
a correspondence between the weak lensing convergence and the underlying density field in the case
of one- and two-point probability distribution function $p(\delta)$ (Munshi
\& Jain 1999b),
\begin{figure}
\begin{center}
{\epsfxsize=7 cm \epsfysize=6 cm {\epsfbox[17 19 457 346]{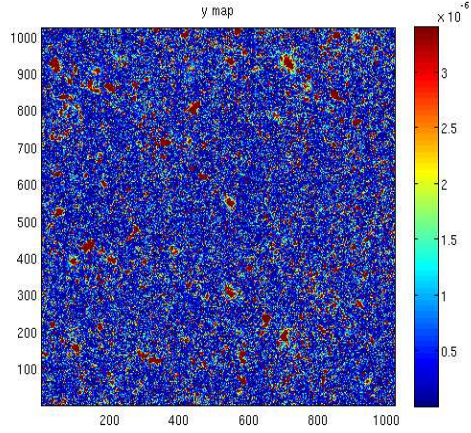}}}
\end{center}
\caption{A simulated map of y-sky used in our study.
The maps are 10 degree on a side and were generated on a $1024 \times 1024$ grid.
These simulations were made using semi-analytical methods developed by \citep{SW03}.
More details
about the simulations can be found in \citep{Wh03}.}
\label{fig:sim1}
\end{figure}
\begin{equation}
\hat b_{}(x_1)\hat h_{}(x_1) \hat b_{} (x_2) \hat h_{} (x_2) =
 b(x_1) h(x_1) b(x_2) h(x_2).
\label{eq:b(x)}
\end{equation}
\n
Where we have used the following relation between $\hat\beta(z)=\tau(z)$
and $\hat b(x)$,
\begin{equation}
\hat b_{}(x) \hat h_{}(x) = -{ 1 \over 2 \pi i}
\int_{-i\infty}^{i\infty} dz \; \hat \tau_{} (z) \; \exp (xz).
\label{eq:bias1}
\end{equation}
\cb{In Eq.(\ref{eq:scaled_vpf}) we have already proved that $\hat h(x)=h(x)$; hence from Eq.(\ref{eq:b(x)}) we deduce that $\hat b(x)=b(x)$.
This means that the bias associated with $\delta \hat y$ is identical to that of the underlying density
constrast $\delta$. This is one of the main result of this paper.}
\n
For all practical purpose we found that the differential bias
as defined above is lot more difficult to measure from numerical
simulations as compared to its
integral counterpart where we concentrate on the bias associated with
peaks above certain threshold,
\begin{equation}
\hat b_{}(>x) \hat h_{}(>x) = -{ 1 \over 2 \pi i}
\int_{-i\infty}^{i\infty} dz \; {\hat \tau_{} (z)\over z} \exp (xz).
\label{eq:bias2}
\end{equation}
\n It is important to notice that although the bias $\hat b_{}(x)$
associated with the tSZ field $\hat y_{s}$ and the underlying
density field are the same, the variance associated with the
density field is very high but the projection effects in the tSZ
field brings down the variance to a value comparable to unity. This
indicates that we can use the integral definition of bias to
recover it from its generating function (see Eq.(\ref{eq:bias1}) and
Eq.(\ref{eq:bias2})). Now, writing down the full two point
probability distribution function for two correlated spots in terms
of the tSZ field $\hat y_{s}(\theta_b)$:
\begin{eqnarray}
&&p(\hat y_1,\hat y_2)d\hat y_1 d\hat y_2 = p(\hat y_1) p(\hat y_2)( 1
+ b(\hat y_1)\xi^{(2)}_{\delta}(r_1,r_2) b_{}(\hat y_2)) d\hat y_1 d\hat y_2;
\quad\quad \hat y_i\equiv \hat y_s(\oh_i).
\end{eqnarray}
%

The results derived here for auto-correlation can be
generalised to cross-correlation analysis; this extension will be presented
elsewhere. It is important to realise that the bias $b(\hat y_{s})$
is related to the bias of ``hot-spots'' in the tSZ-map
and relates their distribution with the overall correlation
structure of the tSZ-maps. The bias $b_{\pi}$ defined earlier on the
other hand relates the 3D pressure distribution $\pi_e({\bf x})$ to
the underlying density distribution $\delta({\bf x})$.
\section{\cbr{Simulations}}
\label{sec:sim}
Two main approaches are generally used to simulate tSZ maps: (i)
semi-analytical methods; (ii) Direct numerical simulations.  In the
following we discuss maps made using both sets of techniques.

\subsection{Simulations generated using semi-analytical Methods}
The semi-analytical methods identifies clusters from purely
collisionless or N-body simulations; simplifying assumptions
regarding the distribution, hydrodynamical and thermodynamic
equilibrium properties of the baryons are then used to create a
baryonic data cube from the underlying dark matter distribution. The
resulting distribution of baryons is then eventually used to create
tSZ $y$ maps. This method was pioneered by \cite{KLT01} and was
later used by \cite{SW03} and \cite{Wh03}. In our study we have made
use of data described in \citep{Wh03}, where more detailed
discussion about these simulations as well as the process of
generating the tSZ y-maps that we have used can be
found\footnote{http://mwhite.berkeley.edu/tSZ/}. The numerical
simulations are rather expensive and semi-analytical methods can
provide reasonably accurate results that can be used to understand
the underlying physical processes which affect the statistics of tSZ
process. The particular cosmology that we will adopt for numerical
study is specified by the following parameter values : 
$\Omega_\Lambda = 0.741, h=0.72, \Omega_b =
0.044, \Omega_{\rm CDM} = 0.215, n_s= 0.964, \sigma_8 = 0.803.$
\subsection{Hydrodynamic Simulations}
Early simulations of the tSZ modeled gravitational collapse and
adiabatic compression but initially ignored the effect of adiabatic
cooling which affects the thermal state of the gas especially in the
halos. It was soon realised, however, that the baryons are subjected
to non-gravitational heating processes - such as the feedback of
energy from supernovae or AGN. Over the years the field of numerical
simulations has matured to such an extent that it is now possible to
include these effects self-consistently and in a reasonably large
simulation box.

\begin{figure}
\begin{center}
\begin{minipage}[b]{0.3\linewidth}
\begin{center}
{\epsfxsize=4.5 cm \epsfysize=4.95 cm {\epsfbox[72 1 431 359]
{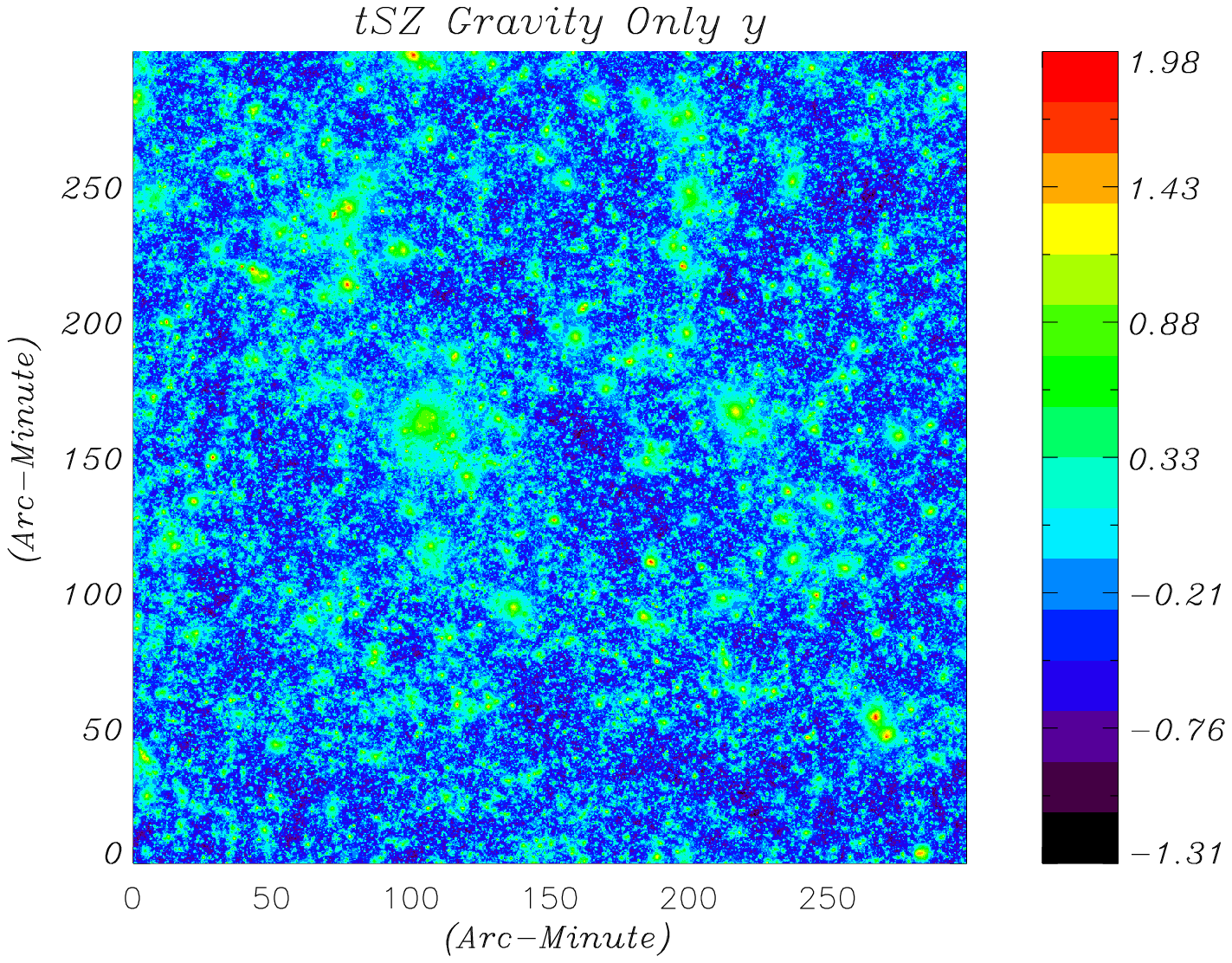}}}
\end{center}
\end{minipage}
\hspace{0.5cm}
\begin{minipage}[b]{0.3\linewidth}
\begin{center}
{\epsfxsize=4.5 cm \epsfysize=4.95 cm {\epsfbox[72 1 431 359]
{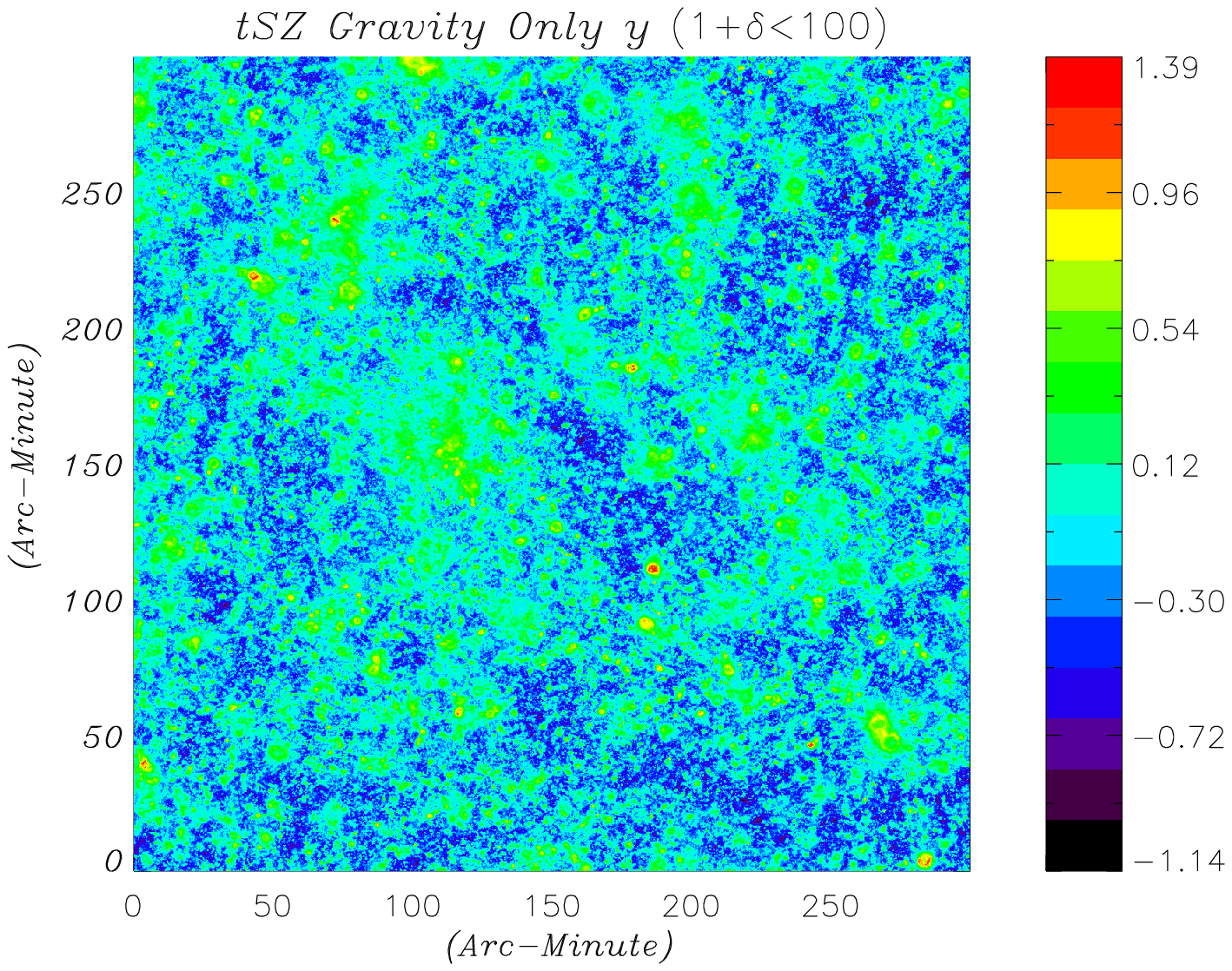}}}
\end{center}
\end{minipage}
\hspace{0.5cm}
\begin{minipage}[b]{0.3\linewidth}
\begin{center}
{\epsfxsize=4.5 cm \epsfysize=4.95 cm {\epsfbox[72 1 431 359]
{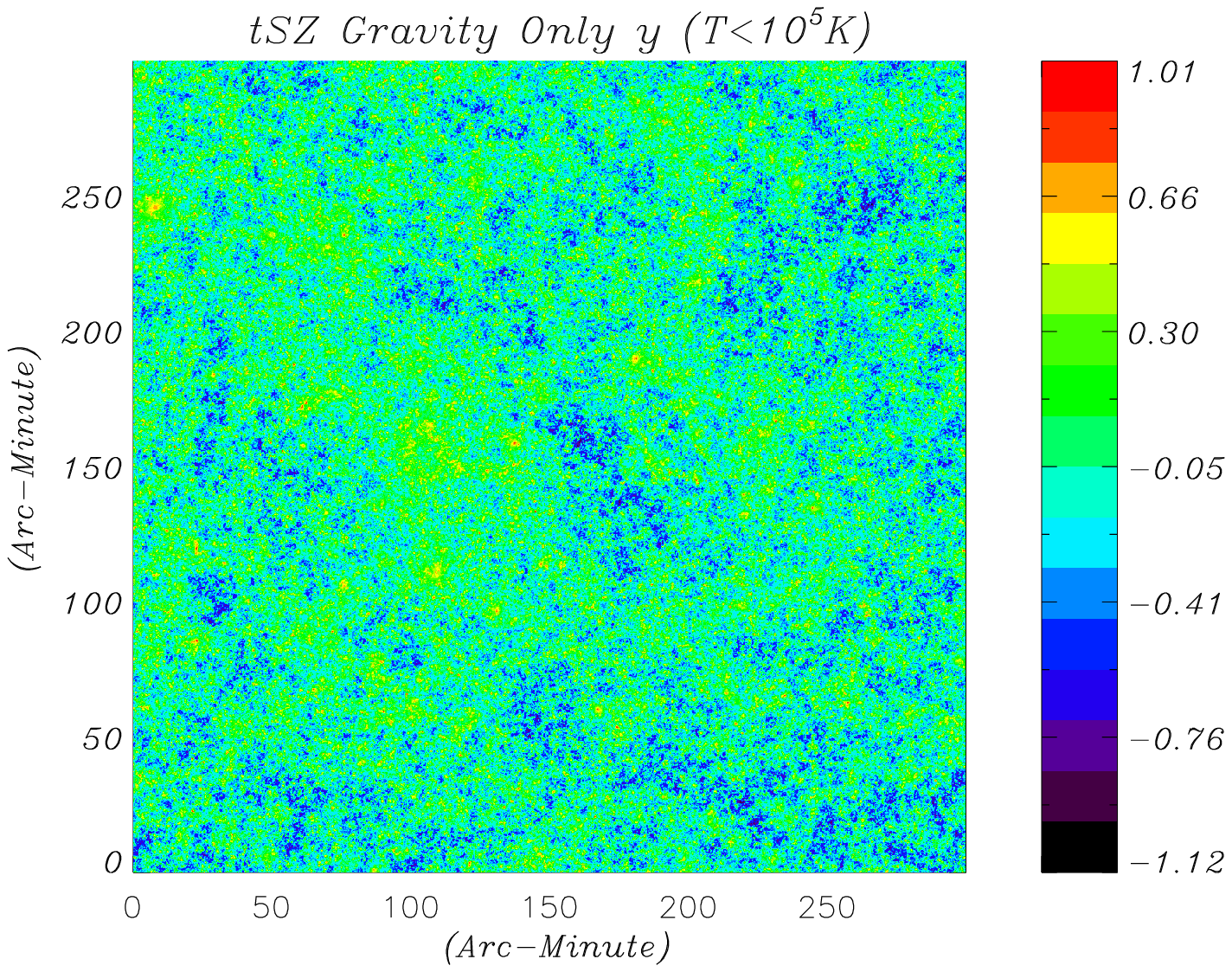}}}
\end{center}
\end{minipage}
\end{center}
\caption{Simulated $5^{\circ}\times 5^{\circ}$ dimensionless
scaled thermal Sunyaev-Zel'dovich maps ${\rm log}_{10}[{y/\la y\ra}]$
are depicted. The maps were generated using Virgo consortium's
Millennium Gas Simulation. The left panel shows the resulting $y$ maps.
The middle panel correspond to
maps generated using low density regions. Only over dense regions
with density $1+\delta <100$ were considered.
The right panel correspond to low temperature regions ${\rm T} < 10^5{\rm K}$.
These set of hydrodynamic simulations ignore pre-heating but takes into account
adiabatic cooling. We will refer to them as GO or Gravity-Only simulations (see text for
more details).}
\label{fig:go}
\end{figure}
\begin{figure}
\begin{center}
\begin{minipage}[b]{0.3\linewidth}
\begin{center}
{\epsfxsize=4.75 cm \epsfysize=4.95 cm {\epsfbox[72 1 431 359]
{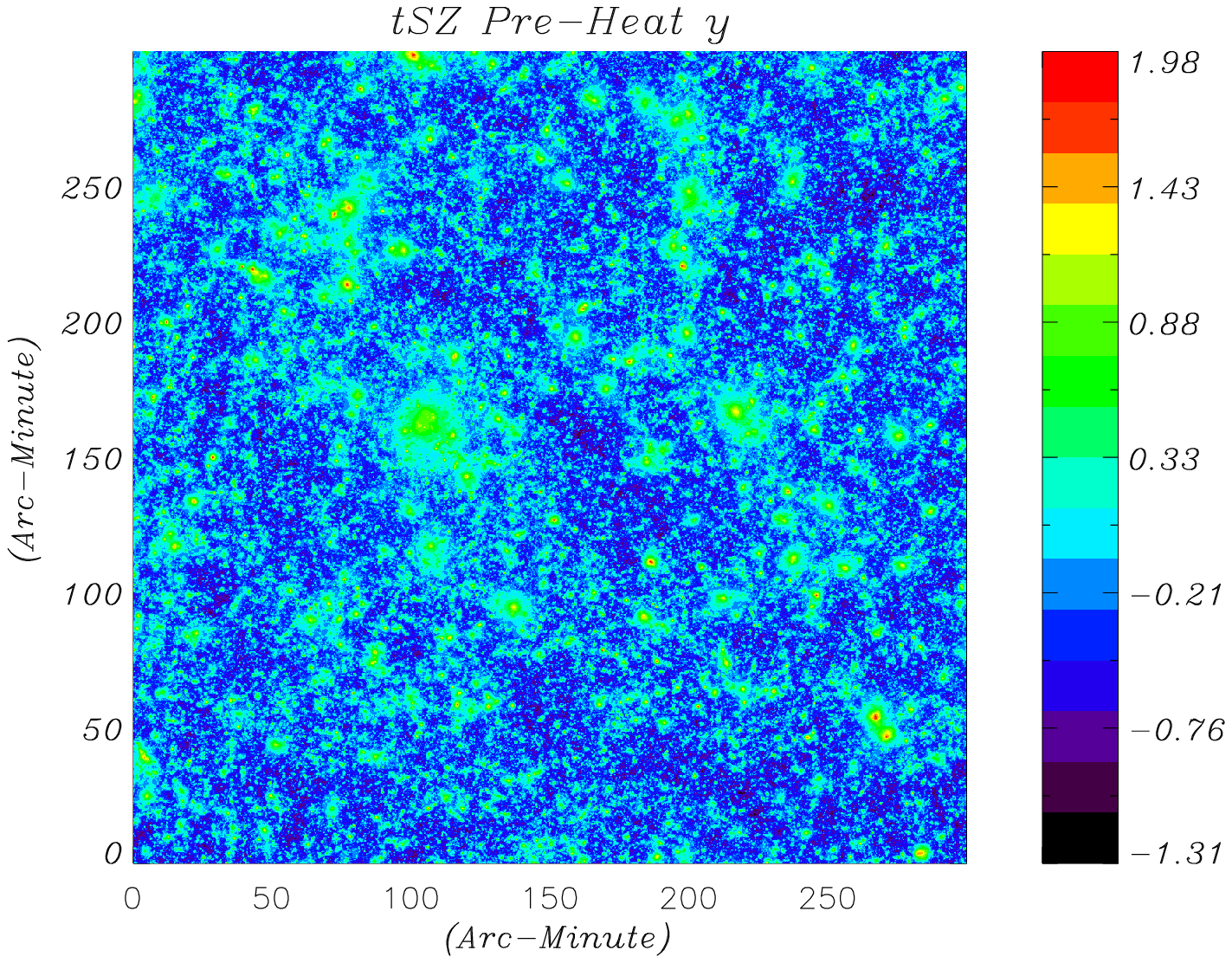}}}
\end{center}
\end{minipage}
\hspace{0.5cm}
\begin{minipage}[b]{0.3\linewidth}
\begin{center}
{\epsfxsize=4.75 cm \epsfysize=4.95 cm {\epsfbox[72 1 431 359]
{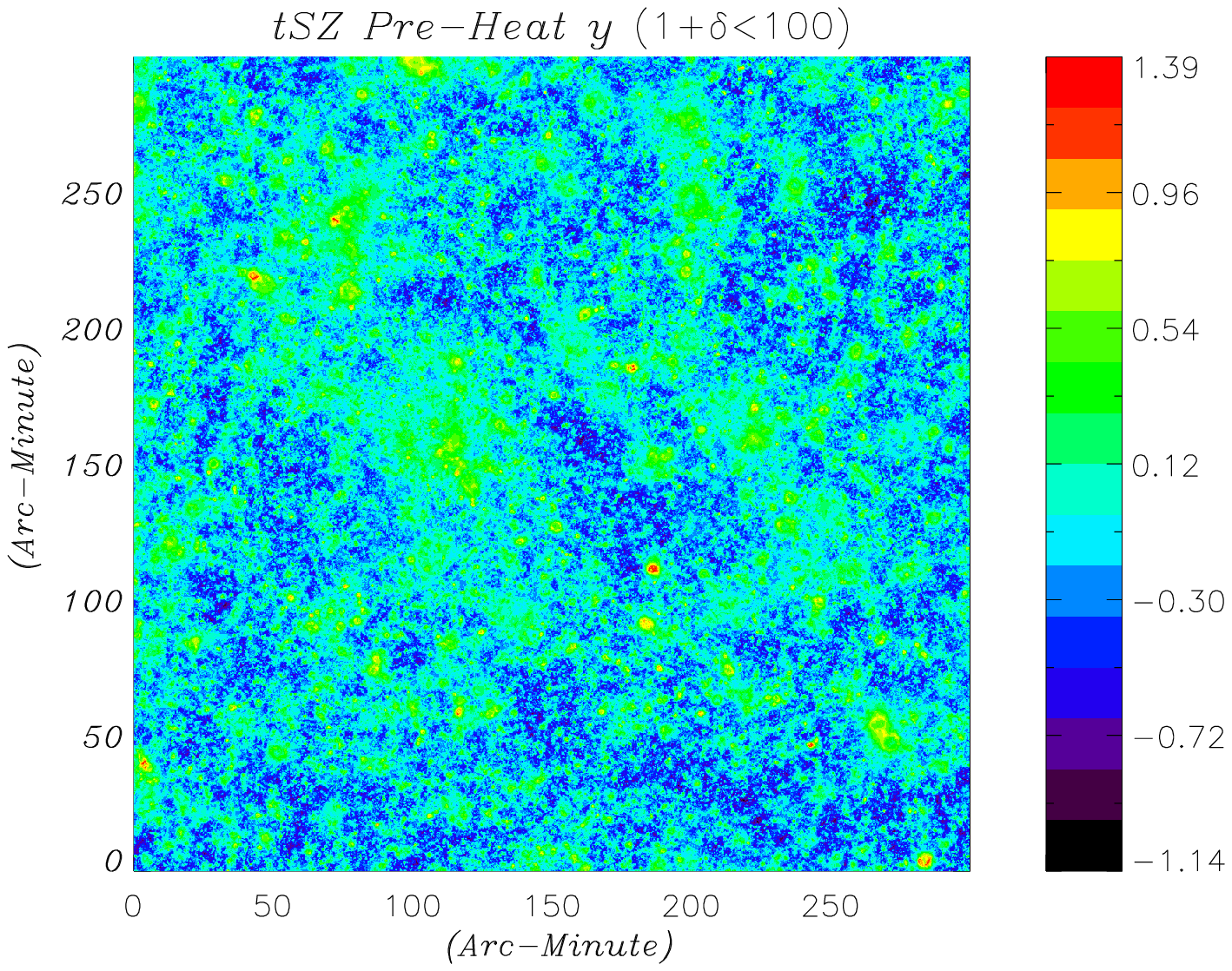}}}
\end{center}
\end{minipage}
\begin{minipage}[b]{0.3\linewidth}
\begin{center}
{\epsfxsize=4.75 cm \epsfysize=4.95 cm {\epsfbox[72 1 431 359]
{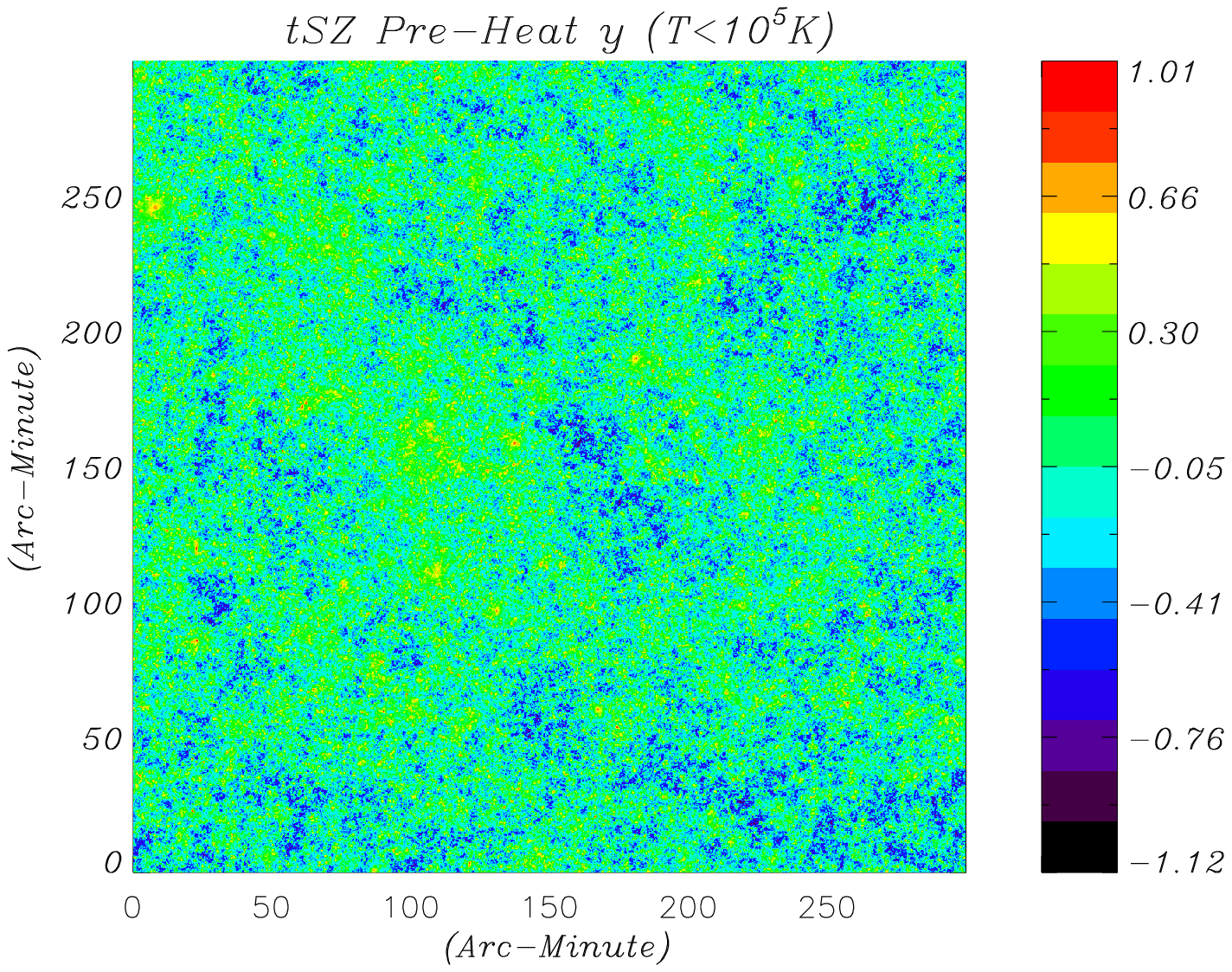}}}
\end{center}
\end{minipage}
\end{center}
\caption{Same as previous figure, but for simulations with pre-heating and cooling.
These simulations will be referred to as PC simulations.}
\label{fig:pc}
\end{figure}

The simulated y-maps that we have used were generated by \cite{Scott12}
using millennium gas simulations  \citep{Hart08,SRE09,Young11, Short10}. Which in turn were
generated to provide hydrodynamic versions of the Virgo consortium's
dark matter Millennium Simulations and were performed using
publicly-available GADGET2 N-body/hydrodynamics code
\citep{Spring05}. Two different versions of the simulations use same
initial conditions and box-size. In the first run, the gas was
modelled as ideal non-radiative fluid and was allowed to go
adiabatic changes in regions of non-zero pressure gradient. The
evolution was modelled using smooth particle hydrodynamics (sph). An
artificial viscosity too was used to convert bulk kinetic energy of
the gas into its internal energy. This is essential to capture the
physics of shock and thus generate quasi-hydrostatic equilibrium.
These process ensures quasi-hydrostatic equilibrium inside
vitalized halos. See text for more details of the hydrodynamic
simulations used to generate these maps. These set of simulations
will be referred as Gravity Only (GO) simulations. Non radiative
descriptions of inter-cluster gas do not reproduce the observed X-ray
properties of the clusters \citep{Voit05}. So the next set of
simulations that we use pre-heated gas at high redshift that can
generate the required core entropy and capable of producing a
steeper X-ray luminosity-temperature in agreement with observations
. The entropy level of these second set of simulations were chosen
to match the mean X-ray luminosity temperature relation at $z=0$
\citep{Kaiser91,EH91}. These simulations also include radiative
cooling and an entropy sink. We will refer to these simulations as
PC. Cooling in these simulations do not play an important role as
the cooling time for the preheated gas is long compared to the
Hubble time. The cosmological parameters of these simulations are
$\Omega_{\rm CDM}=0.25$, $\Omega_{\Lambda}=0.75$, $\Omega_b =0.045$,
$h=0.73$ and $\sigma_8=0.9$.

The scaled  $\log_{10}[{y/\la y \ra}]$ parameter distribution of a
realisation is shown in Figure (\ref{fig:go}) for gravity only or GO
simulations and Figure (\ref{fig:pc}) for simulations with
pre-heating and cooling (PC). The left panels show contribution from
all individual components. The middle panels represents contribution
from over dense regions that satisfy the constraint $1+\delta <
100$. Finally the right panels correspond to the contribution to the
y-map from gas which satisfy the constraint ${\rm T}<10^5{\rm K}$.

There is a very clear and obvious difference between the two sets of
maps in that the GO maps have more substructure. The smoothness of
the PC maps is due to the external thermal energy added to the gas
by the pre-heating process. The mean y-parameter in the GO
simulations is $\la y\ra=2.3\times10^{-6}$ and in the PC simulations
it is nearly four times higher $\la y\ra=9.9\times10^{-6}$. These
values are consistent with COBE/FIRAS constraint $\la y \ra \le 1.5
\times 10^{-5}$. However, it is believed such a high level of
preheating would definitely remove some of the absorption features
seen in the Lyman-$\alpha$ spectrum observed towards quasars \citep{TMS01,SCH07,BV09}. Indeed
the PC model studied here should be treated as an {\em extreme test}
of the effect of a high pre-heating scenario.

In terms of source contributions, the bulk of the $y$-signal comes
from low redshift i.e. $z<2$. However in case of PC simulations the
opposite is true, where $80\%$ of the signal originates from
$z<3.5$. The overdense regions such as the group or clusters are the
sources of $y$-signal in the GO simulations which are primarily
embedded in structures that collapsed at relatively lower redshift.
In case of the PC simulations most of the signals comes from mildly
overdense gas at high redshift. It's interesting to notice that the
GO simulations do get contributions from the gas at high redshift
$z>4$. However such contributions are completely erased in case of
the PC simulations. This is primarily due to the fact that radiative
cooling erases most of the ionized gas at these redshifts.

\section{Tests against Numerical Simulations}
\label{sec:test}
In this section
we present the result of our comparison of theory against the
simulations described in the previous section. We have checked our
results against simulation that are based on N-body simulations
where baryon is added using a semi-analytical prescription. We have
also compared our results against state of the art hydrodynamic
simulations.
\begin{figure}
\begin{center}
{\epsfxsize=15. cm \epsfysize=5. cm {\epsfbox[20 521 588 717]
{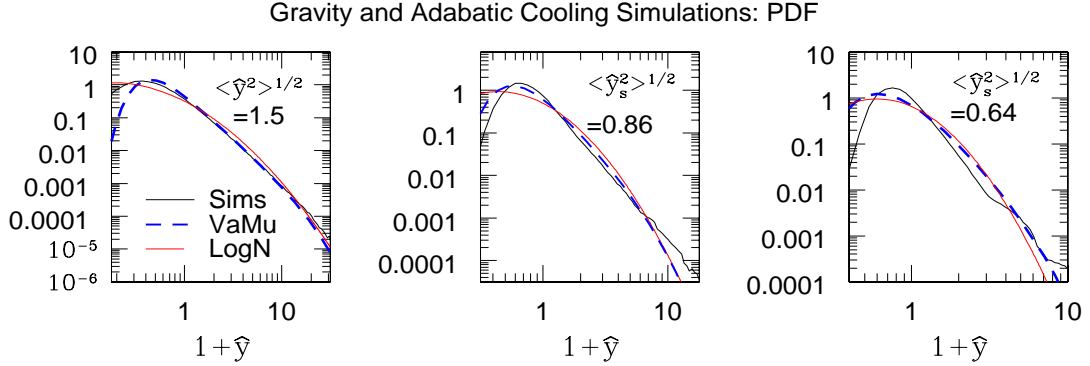}}} \caption{The probability distribution function for
$\hat y$ is being compared to theoretical predictions from lognormal
distribution and an extension of hierarchical model as proposed by
\citep{VaMu04}. The solid-lines represents the PDF computed from the
simulation. The solid and dashed lines correspond to the lognormal
model and the hierarchical model of \citep{VaMu04}. The smoothing
angular scales correspond to $\theta_0 = 0.25''$ (left-panel)
$\theta_0=1.25'$ (middle-panel) and $\theta_0=2.5'$ respectively.
The numerical curves are averages of three individual realisations
each.}
\end{center}
\label{fig:pdf_hydro}
\end{figure}

\subsection{Semi-analytic Simulations}
We have used simulated $y$-maps described in \citep{Wh03} to test our theoretical predictions.
The simulations
were generated on a $1024 \times 1024$ grid and cover $10^\circ \times 10^\circ$
patches on the surface of the sky. To compute the PDF and the bias we construct
the reduced $\hat y_s$ maps from the beam-smoothed $y_s$. A Gaussian
beam with varying FWHM $\theta_b=30'', 1', 5', 10'$ was used for this purpose.
The binning of the data at each grid point was next performed to
the histogram and finally the one-point PDF of $\hat y_s$.
In Figure (\ref{fig:pdf})
we have presented the results of our calculations for the entire range of beam sizes
and compared them against the results from the simulations.

For computation of the bias $b(\hat y_s)$ from the precomputed $\hat
y_s$ we found that the estimation of cumulative bias $b(>\hat y_s)$
is much more stable then its differential counterpart $b(\hat y_s)$.
We used the following expression for computation of the bias
function: \be b(>\hat y_s)= {1 \over \sqrt{\la \hat y_1\hat y_2
\ra}} \left [ {\int_{\hat y_s}^{\infty} d \hat y_1 \; \int_{\hat
y_s}^{\infty} d \hat y_2 p(\hat y_1, \hat y_2) \over [\int_{\hat
y_s}^{\infty} dy_1 p(\hat y_1)]^2} -1 \right ]. \label{eq:bias_est}
\ee The one- and two-point PDFs as well as the two-point correlation
function are all constructed at grid points. We found that the
numerical computation of $b(>\hat y_s)$ to be much more stable than
that of the differential bias $b(\hat y_s)$. The computation of
two-point PDFs were done using different separation angular scales
$\theta_{12}$. The theoretical bias is computed at the large
separation limit. However, we found that this limit is reached very
quickly, typically at an angular scale which is twice the FWHM i.e.
$\theta_{12}\sim 2\theta_0$. We also notice that the estimation of
the bias is independent of any assumption regarding the
factorizability of 2PDF; i.e. Eq.(\ref{eq:bias_est}) do not depend
on any such {\em ad hoc} assumptions.

In Figure (\ref{fig:bias}) we have shown the integrated bias
$b(>\hat y_s)$ computed using two different analytical techniques as
function of the threshold $y_s$. In the right panel the various smoothing angular
scales FWHM we have shown correspond to $\theta_b=30'', 1', 5',
10'$. The solid lines correspond to the numerical simulations.
The dashed lines correspond to predictions from the lognormal model.
This bias for higher $\hat y_s$ regions is higher.
The results for the perturbative calculations are shown only
for $\theta_b=10',5'$. For smaller beam the perturbative
calculations break down.

By construction, the statistics of the $\hat y$ field are
insensitive to the background cosmology, in that they mimic the
statistics of the underlying density contrast $\delta$. Hence the
shape of its PDFs do not depend on the detailed modelling of the
electron pressure $\pi_e$ or the related electron-pressure bias
$b_{\pi}(r)$. The variance of these distributions however do depend
on this function. We have used the variance $\la \hat y_s^2\ra_c$
computed from the maps in our computation of the PDFs.
\begin{figure}
\begin{center}
{\epsfxsize=15. cm \epsfysize=5. cm {\epsfbox[20 521 588 717]
{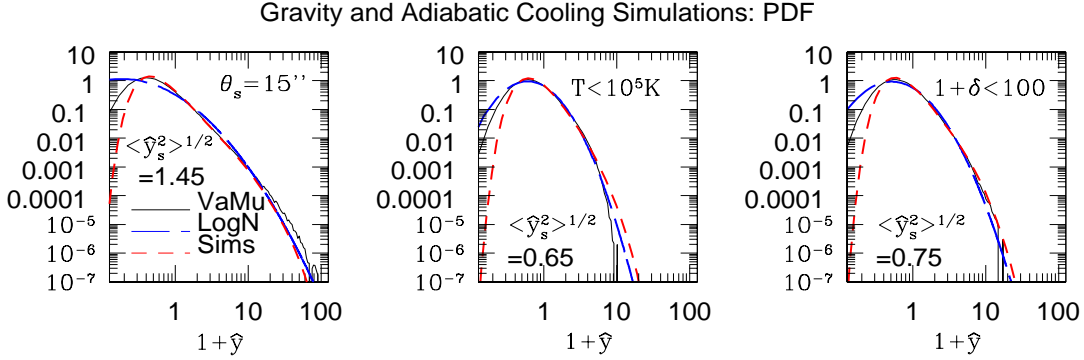}}} \caption{The probability distribution
function for $\hat y$ is being compared to theoretical predictions
from lognormal distribution and an extension of hierarchical model
as proposed by \citep{VaMu04}. The solid-lines represents the PDF
computed from the simulation. The solid and dashed lines correspond
to the lognormal model and the hierarchical model of
\citep{VaMu04}. The smoothing angular scales correspond to $\theta_0
= 0.25''$ (left-panel) $\theta_0=1.25'$ (middle-panel) and
$\theta_0=2.5'$ respectively. The numerical curves are averages of
three individual realisations each.}
\label{fig:no_pdf}
\end{center}
\end{figure}
\begin{figure}
\begin{center}
{\epsfxsize=15. cm \epsfysize=5. cm {\epsfbox[20 521 588 717]
{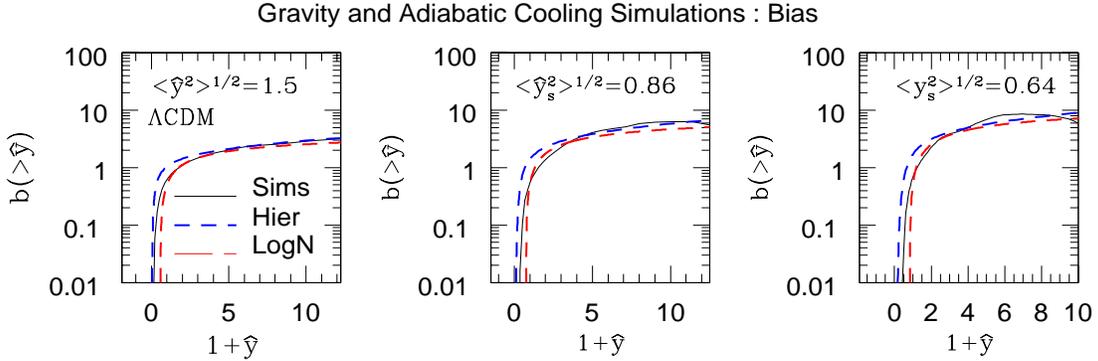}}}
\label{fig:bias1}
\caption{The bias $b(>y_s)$ plotted as a function of $y_s$ for various
smoothing angular scales correspond to $\theta_0 = 0.25''$ (left-panel)
$\theta_0=1.25'$ (middle-panel) and $\theta_0=2.5'$. Theoretical predictions
from hierarchical ansazt (short-dashed), lognormal (long-dashed) and simulations
(solid-lines) are shown.}
\end{center}
\end{figure}
\subsection{Hydrodynamical (SPH) Simulations}
\label{sub:hydro}
In addition to the maps generated using semi-analytical
methods we have also used maps generated by realistic state of the
art hydrodynamic simulations. These maps are $5^\circ\times5^\circ$ in size
and are constructed using a $1024 \times 1024$ grid. We have analysed two different
sets of maps to test our analytical predictions which we describe below.
\subsubsection{Simulations with Adiabatic Cooling:}
\label{sunsubsec:adia}

The first set of maps are derived from the GO simulations described
previously. To these we have compared both the lognormal
approximation and the predictions from analytical results of
\cite{VaMu04}. We have tried three different angular scales
$\theta_b=15'', 1.25', 2.5'$ respectively. We numerically evaluate
the statistics of $\hat y_s$, using exactly the same technique
described before. We find that the results from numerical
simulations are reproduced extremely well in the realistic and
up-to-date hydrodynamic simulations. In case of the GO simulations,
where gravitational dynamics and adiabatic cooling are the main
factors influencing structure formation ,the numerical PDF is reasonably
reproduced by theoretical predictions. We have presented these
results in Figure ({\ref{fig:pdf_hydro}}). As noted before,
perturbative predictions start to break down when the variance at
a given scale reaches unity. The theoretical predictions of
\cite{VaMu04} remedies the situation and is useful for analytical
prediction of PDF at an arbitrary non-linear scale. We find that
analytical prediction for $\hat y_s$ is accurate down to very low
values of PDF. It reinforces our conclusions drawn from analysis of
maps generated using semi-analytical techniques. Though the maps
from hydrodynamical simulations  have less sky-coverage compared to
the semi-analytical maps used before they have more realistic
representation of the baryonic physics responsible for the tSZ
effect.

In addition to studying the $\hat y_s$ maps, we have divided the
entire contribution from various baryonic components, to check how
our theoretical prescriptions compare with that from simulations for
individual components. In this context, we notice that,
thermodynamic states of baryons, as well as their clustering, at low
to medium redshift $z<5$, has been studied, using both numerical as
well as analytical techniques. In their studies, \cite{VSS02} has
used the hierarchical ansatz, to study the {\em phase-diagrams} of
cosmological baryons as function of redshift. The low temperature
``cool'' component of the intergalactic medium (IGM) represented by
Lyman-$\alpha$ forest typically satisfies the constrain $10^3{\rm
K}<{\rm T}<10^4{\rm K}$. The exact values of the lower and
upper-limit depends somewhat on the redshift. The ``cold'' component
of the IGM is very well characterized by a well-defined equation of
state. The ``warm'' component of the IGM on the other hand is
shock-heated to a temperature range of $10^4{\rm K}<{\rm T}<10^7{\rm
K}$ due to the collapse of non-linear structure and can not be
defined by a well defined equation of state. Though the ``warm''
component does follow a mean temperature-density relation, the
scatter around this relation however is more significant than for
the ``cool'' component. Both the ``cool'' and ``warm'' components
originate outside the collapsed halos and typically reside in
moderate overdensites $1+\delta<100$. Finally the remaining
contribution comes from the hot baryonic component of the virialized
high density halos with temperatures ${\rm T}>10^7{\rm K}$.

In  Figure (\ref{fig:no_pdf}) we have compared the contributions
from the medium over-density regions $1+\delta<100$ (medium panel)
which is caused by both ``warm'' and the ``cool'' component of the
IGM. The right panel correspond to emission primarily from the
``cool'' component. The variance corresponding to these individual
components are computed individually and used as an input in the
computation of their PDF. The results are presented for the
particular case of smoothing $\theta_s=15''$ although the agreement
is equally good for other smoothing angular scale. The sharp drops
seen in the PDFs are due to the finite size of the catalog. The
lowest probability that we can compute using a $1024\times 1024$
grid is roughly $10^{-6}$. For larger smoothing angular scales it is
slightly less.

The computation of integrated bias $b(>y_s)$ was carried out using
techniques discussed in the previous section. We find that the
analytical predictions for bias to match very accurately with the
one recovered from numerical simulations. These results are
presented in Figure (\ref{fig:bias}). The bias were estimated for
the same angular scale. Indeed, in almost all cases the predictions
for lognormal as well as hierarchical model are very close. An
extension of \cite{VaMu04} technique for the case of bias is
possible, but has not been worked out. However, bias computed using
hierarchical model and lognormal approximation seems to be
reasonably accurate. The bias $b(>y_s)$ being a two-point statistic
is more sensitive to sample variance.
\subsubsection{Simulations with Adiabatic Cooling and Pre-Heating:}
\label{subsubsec:pree_cool}
\begin{figure}
\begin{center}
{\epsfxsize=15. cm \epsfysize=5. cm {\epsfbox[20 521 588 717]
{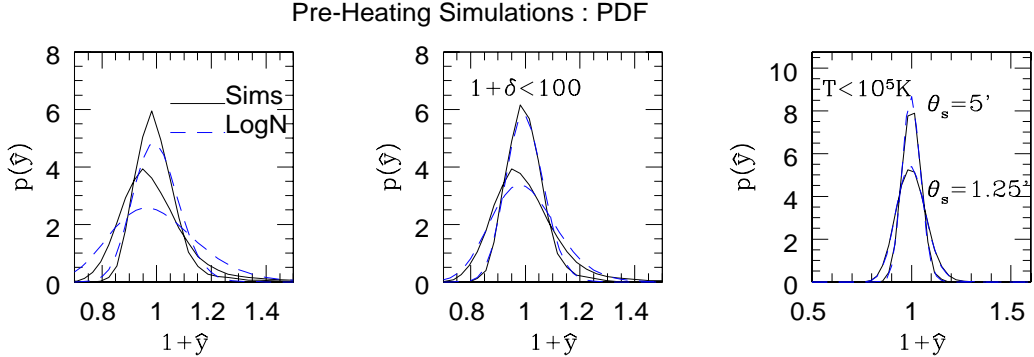}}} \caption{The PDF $p(\hat y)$ of  $\hat y$ is
compared to theoretical predictions from lognormal distribution and
an extension of hierarchical model as proposed by \citep{VaMu04}.
From left to right panels correspond to total contribution,
contribution from regions of low overdensity $1+\delta< 100$ as well
as contribution only from low temperature regions ${\rm T}<10^5 {\rm
K}$ of the simulations. In each panel we show two smoothing scales
$\theta_s=1.25'$ as well as $\theta_s=5'$ respectively.}
\label{fig:pre_heat}
\end{center}
\end{figure}

\begin{figure}
\begin{center}
{\epsfxsize=15. cm \epsfysize=5. cm {\epsfbox[20 521 588 717]
{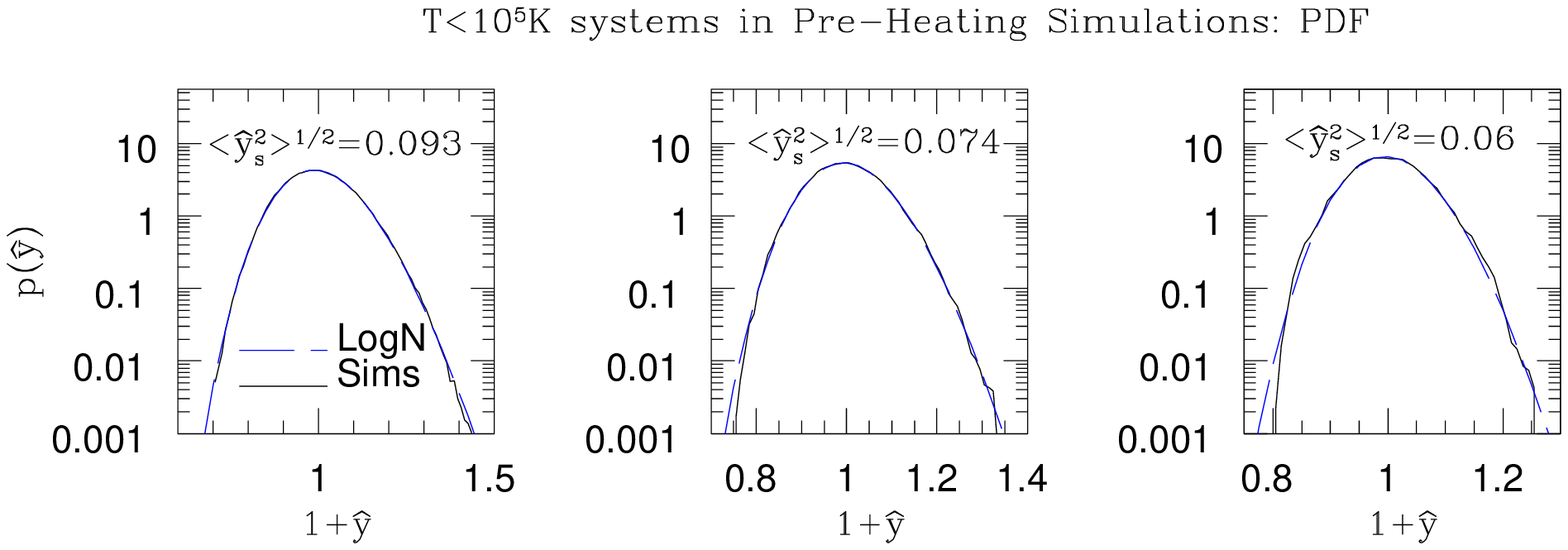}}} \caption{The pdf for $\hat y$, $p(\hat y)$ from
numerical simulations (solid-lines) for ``cold'' ${\rm T}<10^5{\rm
K}$ components are compared against the theoretical predictions from
lognormal distribution (dashed-lines). The hierarchical model as
proposed by \citep{VaMu04} produces nearly identical results and are
not shown. The angular scales considered from left to right are
$\theta_s=0.25''$, $\theta_s=1.25'$ and $\theta_s=2.5'$
respectively. }
\end{center}
\label{fig:lowT}
\end{figure}

The second set of simulations that we have analysed includes
pre-heating and is referred to as the PC simulations. The smoothness
of the PC maps are reflected in their low variance. This is
primarily due to the high level of the pre-heating that erases many
substructures resulting in maps with less features. We include these
simulations in our studies mainly to test the limitations of
analytical predictions. The fundamental assumption in our analytical
modeling is of gravity induced structure formation where baryons are
considered as the biased tracers of underlying dark matter
clustering. We find significant deviation of the numerical results
from theoretical predictions in the presence of high level
preheating at small angular scales Figure (\ref{fig:pre_heat}).
These deviations are more pronounced at smaller angular scales. The
PDFs become Gaussian at scales $\theta_s \sim 10'$ or larger. The
deviation at all-scales is less pronounced if we remove the
collapsed objects and focus on the maps with overdensites $1+\delta
< 100$. However the PDF of the $y$ distribution from ``cold''
intergalactic gas is represented very accurately by our analytical
results at all angular scales in the presence of pre-heating Figure
(\ref{fig:lowT}).
\section{Conclusions}
\label{sec:conclu}
%

We have studied the prospects for extracting and using the
non-Gaussian statistical signatures from tSZ maps. The tSZ effect is
associated with the hot gas in large-scale structure that is probed
by multi-frequency CMB experiments. When compared to the CMB
temperature anisotropies, the tSZ effect has a distinct spectral
dependence with a null at a frequency of 217 GHz. This distinct
spectral signature of SZ effect means it can be effectively
separated from the primary CMB contributions. This will provide an
unique opportunity to probe tSZ effect using data from ongoing
surveys.

%
The statistical analysis of frequency--separated tSZ maps has so far
been mainly focused on lower order statistics or topological
descriptors \citep{Mu12}. Non-Gaussianity in the tSZ signal is an
additional information that is useful in constraining the large
scale pressure fluctuation associated with the tSZ effect. This
signature can be useful for constraining non-gravitational effects
such as pre-heating or other forms of energy injection in the form
feedback from active galactic nuclei (AGN) or super novae (SN).

%
The tSZ effect traces the pressure fluctuations associated with the
large scale distribution of the baryonic gas. The pressure
fluctuations due to the virialized dark matter halos can be modeled
by assuming them to be in hydrostatic equilibrium with the dark
matter distribution in the halo. Such a halo model description has
been extensively used in understanding the statistical properties of
tSZ effect as well as other CMB secondaries \citep{CooSeth02}.

In addition to the halo model prescription, a redshift-dependent
linear biasing scheme that relies on  a perturbative description of
dark matter clustering has also been in use to model certain aspects
of tSZ effect. It was introduced by \cite{GS99a,GS99b}. Combining
such a model with prescriptions from hyper-extended perturbation
theory of \cite{SF99}, can be used to predict the lower-order
moments; this approach has been tested successfully against
numerical simulations by \cite{CHT00}.

In the following we summarize the main conclusions of this study.

\noindent {\bf Use of perturbative approach and its extension by
\cite{VaMu04}}: We have used the approach developed by
\cite{GS99a,GS99b,CHT00} to construct the PDF and bias. We construct
the entire cumulant-generating function and show that under certain
simplifying assumptions it becomes independent of the details of the
biasing scheme. The generating function adopted here was developed
primarily for the construction of 3D and 2D (projected) density
distribution that are studied using galaxy surveys; later it was
used in construction of weak lensing PDF in small and large
smoothing angular scales. In this paper we have shown that a similar
technique can be adopted for the study of tSZ when it is expressed
in terms of the 3D pressure fluctuations. We have compared these
results against two sets of maps: (a) maps made from semi-analytical
simulations which are of $10^{\circ}\times 10^{\circ}$in size (b)
$5^{\circ}\times 5^{\circ}$ maps using full SPH simulations (the
millennium gas simulations from Virgo consortium). We have studied
angular scales $10''<\theta_b<10'$ for the semi-analytical maps and
$2.5''<\theta_b<5'$ for the smaller maps. On angular scales where
the rms fluctuation in $y_s$ maps is lower than unity the
perturbative series is valid and the analytical predictions are in
very good agreement with numerical simulations; we find analytical
results from perturbation theory to be accurate for angular scales
larger than few arc-minutes ($\theta_b>2'$). As a result of our
study we notice that even at comparatively large angular scales,
$\theta_b>10'$, the PDF distribution of the tSZ effect is highly
non-Gaussian. Going beyond the perturbative approach we have used
the formalism presented in \cite{VaMu04} to extend the analytical
predictions to all angular scales. We find an excellent agreement
with theory and simulations for all possible angular scales studied
by us for simulation where gravity and adiabatic cooling plays a
dominant role.
\noindent {\bf Use of lognormal approximation: }In addition to the
perturbative approach and its extension using hierarchical
\emph{ansatz} we have also used a model based on the lognormal
distribution. The lognormal model is non-perturbative and has been
used widely in the literature. Like the perturbative approach it has
been used to model the results from galaxy surveys
\citep{Ham85,CJ91,Bouchet93,Kf94}, weak lensing observables
\citep{Mu00,TTHF02} as well as Lyman-alpha statistics \citep{BD97}.
The validity of the lognormal model has been compared against
numerical simulations as well as against perturbative or
hierarchical methods \citep{BK95}. In the entire range of  angular
scales considered by us we find the lognormal model to be a very
good approximation for the $\hat y$ parameter distribution. We have
also used the bias computed from lognormal approximation and found
it to be in reasonable agreement.

\noindent{\bf Isolating the Effect of Background Cosmology:} We have
shown that the statistics of $\hat y$ can be described using
analytical models of gravitational clustering alone and the $S_N$
parameters that describe the PDF as well as the entire PDF is
insensitive to the background cosmology - the lognormal model e.g. do
not have any cosmology built into it. The power spectrum of $y$ on
the other hand is sensitive to the amplitude of the density
fluctuations, $\sigma_8$ and other cosmological parameters. Using
higher order statistics, such as the skewness, previous authors
argued the possibility of separation of the pressure bias from the
amplitude of the density fluctuations \citep{HS12}. The dimensionless parameter
$\hat y$ that we have introduced here is insensitive to background
cosmology and in this sense our approach achieves separation of
background cosmology and the effects of gravitation to all order.

\noindent{\bf Separation of gravitational and non-Gravitational
Aspects:} Though the PDF constructed using the semi-analytical
approaches do agree with numerical simulation that incorporates
gravitation and adiabatic cooling, we also find departure from
analytical prediction in case of simulations with pre-heating. The
non-gravitational processes that include pre-heating is not included
in our analytical calculations. Thus departure from theoretical
predictions provides a particularly interesting approach to separate
out the effects of non-gravitational process on tSZ statistics.

\noindent{\bf Non-Gaussianity contribution from different baryonic
components:} We have studied the contribution to the tSZ effect from
individual components such as the ``cold'' gas $\rm T<10^5\rm K$,
uncollapsed moderate overdense ``warm'' gas $1+\delta<100$ and the
total contribution from all components that include the
intra-cluster gas within the halos. We find that in addition to the
total tSZ maps, the individual tSZ maps constructed using  $\rm
T<10^5\rm K$, $1+\delta<100$ components too can be described using
our approach. For simulations the with pre-heating we find a
significant departure for the
 contribution from ``warm'' gas  component $1+\delta<100$. The departure is
more pronounced when the contributions from virialised halos are
included. However we also find a near-perfect match for the
contribution from the cold component.

Many previous studies have considered a halo model based approach
for modeling of the gas distribution in collapsed, virialized
objects. In this approach, the specific number density, and the
radial profile of these halos are analyzed using a Press-Schecther
formalism or its variants. However, detailed modelling of the
complete PDF or the bias is possible in this approach only in an
order by order manner. In our extended perturbative approach or the
lognormal analysis, we go beyond the order-by-order approach and
construct the entire cumulant generating function $\Phi(z)$ and
relate it to that of the underlying density distribution $\phi(z)$.
This allows us to reproduce and predict the entire PDF of the tSZ
distribution for a specific smoothing angular scale. The statistical
picture that we have developed here is complementary to that based
on the Press-Schechter formalism.

We would also like to point out that some previous studies have
modeled the statistics of tSZ effect using the hierarchical {\em
ansatz} \citep{VS99,VSS02}. However in these studies the
contribution from individual halos were computed using a
virialization scheme and equilibrium profile for the pressure
distribution within the halos. The results presented here are
complementary to such prescriptions as we focus on the large-scale
distribution of ionized gas. Instead of modeling the contribution
from individual halos, we directly link the baryonic pressure
fluctuation responsible for the large-scale tSZ effect in terms of
the underlying mass distribution.

We have ignored the presence of instrumental noise in our results. A
beam-smoothed noise of known PDF (Gaussian or otherwise) can always
be convolved with the theoretical PDFs presented here before
comparing them with any observational data. The method we pursue
here also relies on having access to frequency cleaned tSZ maps. The
tSZ effect can also be studied using cross-correlation techniques
that involve external tracers; such methods typically employ the
``mixed'' bispectrum. The results, however, lack frequency
information and are typically dominated by confusion noise. Using
frequency cleaned maps is expected to enhance the signal-to-noise
significantly by exploiting frequency information, in the absence of
which the background CMB plays the role of intrinsic noise that
degrades the signal-to-noise ratio. It is also interesting to note
that the removal of tSZ signals from the CMB maps may actually help
detection of other sub-dominant effects. The study of PDFs presented
here can play important role in this direction.
\section{Acknowledgements}
\label{acknow}
DM and PC acknowledge support
from STFC standard grant ST/G002231/1 at School of Physics and
Astronomy at Cardiff University where this work was completed.
We would like to thank Alan Heavens, Patrick Valageas, Ludo van Waerbeke and Martin White for many useful discussions.
We would like to thank Martin White for making his tSZ simulations data freely available which we have used in this work. 
We would also like to thank Francis Bernardeau for making a copy of his code available to us which we have modified to
compute the PDF and bias of the tSZ field for the perturbative model. SJ and JS acknowledge support from from the US
Department of Education through GAANN fellowships at UCI. We would also like to acknowledge many useful suggestions from our referee who
helped us to improve the draft of this paper. 
\bibliography{paper.bbl}
\appendix
\section{The Lognormal Distribution}
\begin{figure}
\begin{center}
{\epsfxsize=10 cm \epsfysize=5 cm {\epsfbox[27 426 588 709]{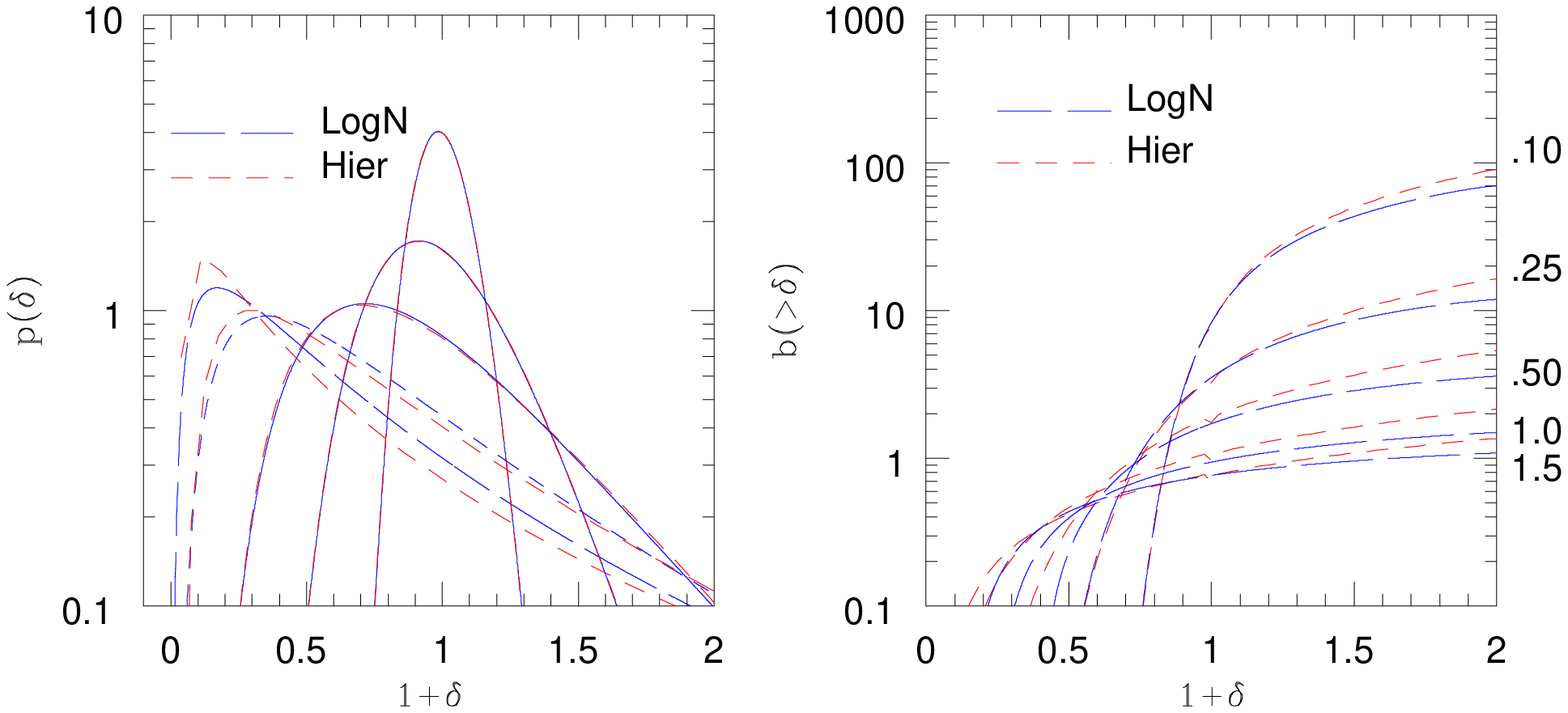}}}
\end{center}
\caption{The left panel depicts PDF $p(\delta)$ as a function $1+\delta$
for various values of $\sigma =0.1, 0.25,0.5,1.0, 1.5$. Two different approximations are considered,
the lognormal (long dashed) and the hierarchical ansatz (short dashed).
The broader PDFs correspond to higher values of $\sigma$.
The bias $b(>\delta)$ is plotted as a function $1+\delta$ in the right panel. Two different approximations are considered,
the lognormal distribution (long dashed) and the hierarchical ansatz (short dashed). The analytical results
correspond to the large separation limit.}
\label{fig:pdf_lgn}
\end{figure}
The evolution of PDF of the density field $\delta$ can also be
modelled using {\em lognormal} distribution.
\citep{Ham85,CJ91,Bouchet93,Kf94}. Detailed discussion of various
issues involving of lognormal distribution can be found in
\cite{BK95,Co94}. We will use the following expressions for the PDF
$p(\delta)$ and the joint-PDF $p(\delta_1,\delta_2)$ for our study
\citep{TTHF02}:
\begin{eqnarray}
&& p(\delta)d\delta = {1 \over \sqrt {2\pi \Sigma^2}} \exp \left [ -{\Lambda^2 \over 2\Sigma^2}\right ]{d\delta \over 1+\delta};
\label{eq:logn1}\\
&& \Sigma_{}=\ln(1+\sigma^2); \quad \Lambda = \ln[(1+\delta)\sqrt{(1+\sigma^2)}\cb ];\label{eq:logn1a} \\
&& p(\delta_1,\delta_2)d\delta_1 d\delta_2 =  {1 \over 2\pi \sqrt {\Sigma^2 - X_{12}^2}}\exp \left [ -{\Sigma(\Lambda_1^2 + \Lambda_2^2) -2X_{12}\Lambda_1\Lambda_2 \over 2(\Sigma^2 - X_{12}^2)}\right ] {d\delta_1\over 1+\delta_1} {d\delta_2\over 1+\delta_2}; \label{eq:logn2}\\
&& \Lambda_i = \ln[(1+\delta_i)\sqrt{(1+\sigma^2)} \cb];  \quad X_{12}=\ln(1+\xi^{(2)}_{\delta}(r_1,r_2)).
\label{eq:logn2a}
\end{eqnarray}
\n
In the limiting case of large separation $X_{12}\rightarrow 0$ we can write down the two point PDF as:
\be
p(\delta_1,\delta_2)= p(\delta_1)p(\delta_2)[1+ b(\delta_1)\xi_{\delta}^{(2)}(r_1,r_2)b(\delta_2)];
\quad\quad  b(\delta_i)= \Lambda_i/\Sigma_{}.
\ee
\n
It is however easier to estimate the cumulative or integrated bias associated with objects beyond a certain density threshold $\delta_0$.
This is defined as $ b(\delta>\delta_0)=\int_{\delta_0}^{\infty} p(\delta) b(\delta) d\delta / \int_{\delta_0}^{\infty} p(\delta) d\delta$.
In the low variance limit $\sigma^2 \rightarrow 0$ the usual Gaussian result is
restored $b(\delta)= \delta/\sigma^2$. The parameters $\Lambda,\Lambda_i, X_{12}, \Sigma$ that
we have introduced above can be expressed in terms of the two-point (non-linear) correlation function
$\xi^{(2)}_{\delta}(\br_1,\br_2) \equiv \la\delta({\bf r}_1)\delta({\bf r}_2)\ra \equiv \la \delta_1\delta_2 \ra$ and the nonlinear variance $\sigma^2 = \la \delta^2 \ra$ of the smoothed density field.
Lognormal distribution has already been used to model
the statistics of weak lensing observables \citep{Mu00,TTHF02} and the clustering of Lyman alpha
absorption systems (e.g. \cite{BD97}). and is known to model gravitational clustering
in the quasilinear regime  \citep{MSS94,MLMS92}. In Figure (\ref{fig:pdf_lgn}) we compare the PDF and bias
predictions from the lognormal model and hierarchical ansatz for various values of the variance.
Though hierarchical ansatz and the lognormal model both predict nearly identical PDFs
in the quasilinear
regime, it is important to realize that the lognormal model is {\em not} a member
of the family of hierarchical models, i.e. the PDF of lognormal model can't be cast
into the form given in Eq.(\ref{eq:pdf_scaling}).
\section{Hierarchical Ansatz (Minimal Tree Model): A Very Brief Review}

\begin{figure}
\begin{center}
{\epsfxsize=5 cm \epsfysize=5 cm {\epsfbox[33 259 588 709]{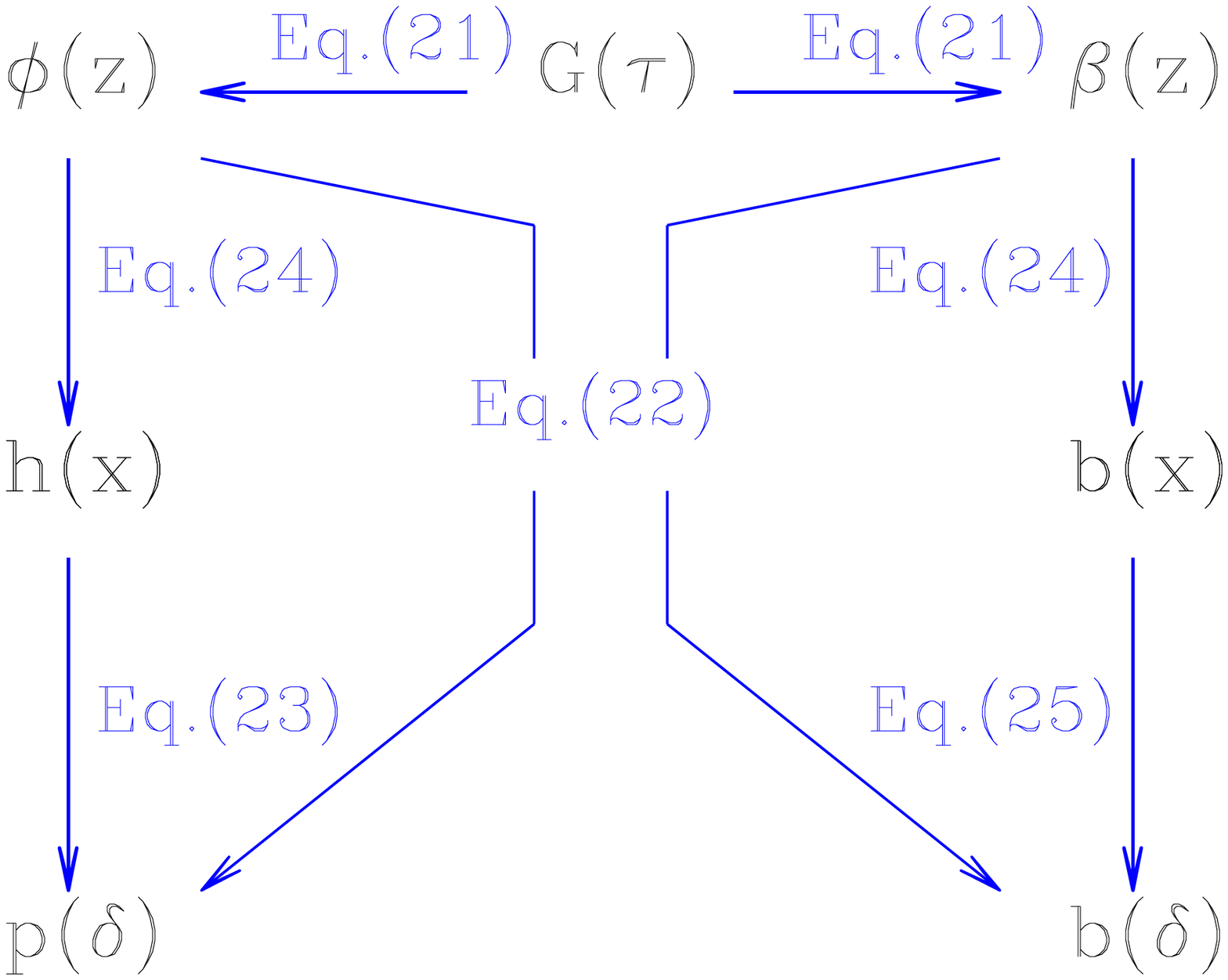}}}
\end{center}
\caption{A simple ``flowchart'' for a hierarchical approach is shown. It provides a summary of how the PDF $p(\delta)$ and bias $b(\delta)$ are
constructed for a given model for the generating function for ${\cal G}(\tau)$. The construction involves the generating function for the
normalised cumulants or $S_p$ parameters and cumulant correlators $C_{pq}$ i.e. $\phi(z)$ and $\beta(z)$. These are
related to $p(\delta)$ and $b(\delta)$ through Laplace transforms. The scaling functions $h(x)$ and $b(x)$ encodes
the scaling properties of $p(\delta)$ and $b(\delta)$. We show the equation numbers that relate various quantities in the diagrams.
The left hand side of the diagram correspond to one-point PDF and the right hand side correspond to the two-point PDF
which is expressed through the bias function $b(\delta)$. }
\label{fig:flowchart}
\end{figure}

As mentioned before the PDF $p(\delta)$ and the bias $b(\delta)$ can both be constructed from the knowledge of the VPF $\phi(z)=\sum_{p=1} S_p {z^p/p!}$
and its two-point analog $\tau(z) =\sum_{p}C_{p1} {z^p/p!}$. Where the parameters $S_p$ and $C_{p1}$ are normalized
cumulants and cumulant correlators for the density field.

The modelling of $\phi(z)$ and $\tau(z)$ needs a detailed knowledge of
the entire correlation hierarchy. The detailed knowledge of the entire correlation hierarchy is
encoded in the vertex generating function ${\cal G}(\tau))$. Typically for large values of
y the VPF exhibits a power law $\phi(z)=az^{1-\omega}$. The parameter typically takes a value
$\omega=.3$ for CDM like spectra. There are no theoretical estimates of $\omega$ and it is
generally estimated from numerical simulations. For small but negative values of $y$
the functions $\phi(z)$ and $\tau(z)$ develops a singularity in the complex plane which
is described by the following parametrization \citep{BS89}:
\be
\phi(z) = \phi_s - a_s \Gamma(\omega_s) ( z - z_s)^{-\omega_s}; \quad\quad
\tau(z) = \tau_s - b_s ( z - z_s )^{-\omega_s - 1}.
\ee
The singularity structure of $\phi(z)$ and $\tau(z)$ depends on the nature of the vertex generating
function $G(\tau)$ and its behaviour near the singularity $\tau_s$ \citep{BS89}:
\be
a_s = {1 \over \Gamma(-1/2)}{\cal G}'(\tau_s) {\cal G}''(\tau_s) \left [
{ 2 {\cal G}'(\tau_s) {\cal G}''(\tau_s) \over {\cal G}'''(\tau_s)}
\right ]^{3/2}; \quad\quad
b_s = \left [
{ 2 {\cal G}'(\tau_s) {\cal G}''(\tau_s) \over {\cal G}'''(\tau_s)}
\right ]^{1/2}.
\ee
On the other hand the parameters $\omega$ and $a$ can be described in terms of a parameter $z_s$
which in turn describes the exponential decay of the PDF for large density contrast $\delta$ \citep{BS89}:
\be
\omega = {k_a \over ( k_a + 2)},\label{ka}; \quad\quad
a = {k_a + 2 \over 2} k_a^{ k_a /  k_a + 2}; \quad\quad
-{ 1 \over z_s} = x_{\star} = {1 \over k_a } { (k_a + 2)^{k_a + 2} \over (k_a + 1)^{k_a+1}}.
\ee
The PDF and the bias thus has two distinct regimes that are dictated by the two asymptotes.
For intermediate values of $\delta$ the PDF shows a power law behaviour. The PDF
and the bias are given by the following expression \citep{BS89}:
\begin{equation}
{\bar \xi_2 }^{ - \omega \over ( 1 - \omega)} << 1 + \delta << \bar \xi_2;
~~~~~~
p(\delta) = { a \over \bar \xi_2^2} { 1- \omega \over \Gamma(\omega)}
\Big ( { 1 + \delta \over \bar \xi_2 } \Big )^{\omega - 2}; ~~~~~
 b(\delta) = \left ( {\omega \over 2a } \right )^{1/2} { \Gamma
(\omega) \over \Gamma [ { 1\over 2} ( 1 + \omega ) ] } \left( { 1 +
\delta \over \bar \xi_2} \right)^{(1 - \omega)/2}.
\end{equation}
For large values of $\delta$ the PDF on the other hand shows an exponential
behaviour \citep{BS89}:
\begin{equation}
1+ \delta >> {\bar \xi}_2; ~~~~
p(\delta) = { a_s \over \bar \xi_2^2 } \Big ( { 1 + \delta \over \bar
\xi_2}  \Big ) \exp \Big ( - { 1 + \delta \over x_{\star} \bar \xi_2}
\Big );  ~~~~~ b(\delta) = -{ 1 \over {\cal G}'(\tau_s)} {(1 + \delta)
\over { {\bar \xi}_2}}.
\end{equation}
At very small values of $\delta$ the PDF shows an exponential decay which
is described only by the parameter $\omega$ \citep{BS89}:
\ben
&& 1 + \delta << \bar \xi_2;~~
p(\delta) = a^{ -1/(1 - \omega)} {\bar \xi}_2^{ \omega/( 1 -
\omega )} \sqrt { ( 1 - \omega )^{ 1/\omega } \over 2 \pi \omega v^{(1
+ \omega)/ \omega } } \exp \Big [ - \omega \Big ( {v \over 1 - \omega}
\Big )^{- {{(1 - \omega)}/\omega}} \Big ]; \\
&& b(\delta) = -\left ( {2 \omega \over \bar{ \xi}_2} \right )^{1/2} \left ({ 1 -
\omega \over v}  \right )^{(1 - \omega)/2 \omega}; \quad
v = (1+\delta)[a]^{[-1/(1-\omega)]} [\bar\xi_2]^{[\omega/(1-\omega)]}.
\een
The range of $\delta$ for which the power law regime is valid depends on
the value of $\bar\xi_2$. For smaller values of $\bar\xi_2$ the power law
regime is less pronounced. A more detailed discussion of these issues can be
found in \cite{MunBer99}. The links to the gravitational dynamics in the quasilinear regime,
for various approximations are discussed in \cite{MSS94}. In this paper we have considered
only a specific version of the hierarchical ansatz known as the {\it minimal} tree model.
This is the most popular version due to its simplicity; for variations
and possible generalizations of minimal tree model see \cite{BS99}.
%
%
%
%
%
%
%
%
%
%
\end{document}